\documentclass[twocolumn]{aastex631}

\shorttitle{Comparison of solar multifrequency microwave data with other solar indices}
\shortauthors{Shimojo et al.}

\begin{document}

\title{Comparison of solar multifrequency microwave data with other solar indices \\ for understanding solar and stellar microwave data}

\correspondingauthor{Masumi Shimojo}
\email{masumi.shimojo@nao.ac.jp}

\author[0000-0002-2350-3749]{Masumi Shimojo}
\affiliation{National Astronomical Observatory of Japan, 
Mitaka, Tokyo, 181-8588, Japan}
\affiliation{Graduate University of Advanced Studies, SOKENDAI, Mitaka, Tokyo 181-8588, Japan}

\author[0000-0002-1297-9485]{Kosuke Namekata}
\affiliation{National Astronomical Observatory of Japan, 
Mitaka, Tokyo, 181-8588, Japan}
 
\author[0000-0002-2464-5212]{Kazumasa Iwai}
\affiliation{Institute for Space-Earth Environmental Research (ISEE), Nagoya University, Nagoya, Aichi, 464-8601, Japan}

\author[0000-0002-5279-686X]{Ayumi Asai}
\affiliation{Astronomical Observatories, Kyoto University, Sakyo, Kyoto, 606-8502, Japan}

\author[0000-0003-0321-7881]{Kyoko Watanabe}
\affiliation{National Defense Academy of Japan, 
Yokosuka, Kanagawa, 239-8686, Japan}

\begin{abstract}
Thermal microwave emissions detected from stellar atmospheres contain information on stellar activity. However, even for the Sun, the relationship between multifrequency microwave data and other activity indices remains unclear. We investigated the relationships among the thermal microwave fluxes with 1, 2, 3.75 and 9.4 GHz, their circular polarizations, and several activity indices recorded during recent solar cycles and observed that these relationships can be categorized into two groups. In the first group, the relationship between the microwave fluxes and solar indices, which are strongly related to the active regions, can be well-fitted by using a linear function. In the second group, the fitting function is dependent on frequency. Specifically, the microwave fluxes at 1 and 2 GHz can be well-fitted to the total unsigned magnetic and extreme ultraviolet fluxes by employing a power-law function. The trend changes around 3.75 GHz, and of the trend for the 9.4 GHz fluxes can be fitted by using a linear function. For the first time, we present the relationship between circular polarization and solar indices. Moreover, we extrapolated these relationships of the solar microwave fluxes to higher values and compared them with the solar-type stars. We found that $\epsilon$ Eri, whose microwave emission originates from thermal plasma, follows the extrapolated relationship. However, to date, only one star's emission at 1 -- 10 GHz has been confirmed as thermal emission. More solar-type stars should be observed with future radio interferometers to confirm that relationships based on solar data can be applied to stellar microwave data. 
\end{abstract}


\section{Introduction} \label{sec:intro} 

The activity of the central star is crucial to the habitability of an exoplanet. Stellar flares and wind, X-rays, and ultraviolet (UV) emissions influence an exoplanet's space weather/climate, atmosphere, and chemistry \citep[e.g.,][]{2019_LinskyBook,2020IJAsB..19..136A}. The knowledge on our Sun reveals that these phenomena strongly depend on the magnetism of the central star. Therefore, the stellar magnetic activity should be studied to understand the effects of the central star on the habitability of an exoplanet. 

Radio observations reveal the behavior of nonthermal electrons in stellar flares and the properties of stellar atmospheres and magnetic fields \citep{1985ARA&A..23..169D}. For example, the total solar flux\footnote{We have observed the total solar "flux density" practically. It is commonly, and perhaps inaccurately, referred to as "flux" in the literature. In this paper, we will use the common term "flux" rather than "flux density".} at 2.8 GHz \citep[F10.7 index:][]{F10.Org, 2013SpWea..11..394T}  is a useful proxy  for the solar extreme ultraviolet (EUV) flux (wavelength: 10 -- 121 nm) that affects the Earth's upper atmosphere  \citep[e.g.,][]{2018SpWea..16..434Z} because microwave\footnote{In this paper, microwave means roughly that its frequency is 1 - 30 GHz. In other words, its wavelength range is 1 - 30 cm.} and EUV radiations originate mainly from the same thermal plasma between the chromosphere and corona. This is a powerful advantage of radio observations, considering that stellar EUV observation is difficult because of absorption by the interstellar medium. Therefore, stellar radio observations are useful for understanding the stellar magnetic activity and are another suitable tool for investigating stellar magnetism, independent of X-ray, far-UV, and optical observations. 

To date, radio-observation-based investigations on stellar phenomena have been rarely conducted, except for stellar flares, because most of the detected radio emissions from stars are considered to originate from stellar flares. Radio emissions from stellar flares have been successfully detected, and the universality of solar and stellar flares has been revealed  \citep[e.g.,][]{1993ApJ...405L..63G, 2002ARA&A..40..217G}. 
Previously, a radio burst with spectral properties similar to those of a solar Type IV burst was detected from Proxima Centauri using the Australian Square Kilometer Array Pathfinder \citep{2020ApJ...905...23Z}, and these results further validate the universality of solar and stellar flares. The radio emissions from flares and radio bursts indicate the existence of nonthermal electrons, which are accelerated by transient phenomena related to stellar activity, such as stellar flares and Coronal Mass Ejections (CMEs). Hence, the signals from stellar flares cannot directly reveal the long-term variations and global properties of stars, e.g., the long-term variation of starspots  \citep{2005LRSP....2....8B,2019ApJ...871..187N,2020ApJ...891..103N}, total magnetic flux \citep{2014MNRAS.441.2361V, 2003ApJ...598.1387P,2022ApJ...927..179T}, and stable components of UV emissions \citep{2019_LinskyBook}, which are crucial for understanding stellar magnetism and its impacts on exoplanets. 

The thermal plasma in the atmospheres of  solar-type\footnote{In this paper, we follow \cite{1998saco.conf...41S}  and define solar-type as the main-sequence stars with 0.50 to 1.00 in B-V ratio. Using the MK classification, they are in the range from F8V to K2V.} stars emits microwaves via two emission mechanisms \citep[e.g.,][]{1985ARA&A..23..169D}. One is the thermal free-free emission (Bremsstrahlung), which originates from all the layers of the atmosphere; however, the dominant component is the radio photosphere at each frequency. In the solar atmosphere, the altitude of the radio photosphere decreases with increasing frequency, and the altitude range for microwaves corresponds to the chromosphere and transition region. Hence, the brightness temperature increases with decreasing observation frequency, and the microwave spectrum does not match that of the optical-thick isothermal plasma \citep{1991ApJ...370..779Z}. Although the optical thicknesses of the ordinary and extraordinary modes (o-mode and x-mode, respectively) for thermal free-free emission depend on the magnetic field \citep{2004ASSL..314...71G}, the circular-polarization degree of this emission from the entire solar/stellar atmosphere is small because the emission primarily originates from the radio photosphere. The magnetic field strength in a corona can be estimated based on its polarization degree by separating the coronal microwave component from the optically thin medium at high frequencies \citep{2016ApJ...818....8M}.

The other microwave emission mechanism is gyroresonance.  \cite{1997SoPh..174...31W} revaluated the emission mechanism of the Sun both theoretically and observationally. The critical property of this emission is that the opacity increases significantly when the thermal electrons gyrated by the magnetic fields resonate with the electromagnetic waves. The opacity caused by the resonance in the x-mode is significantly larger than that in the o-mode. Hence, the gyroresonance emission is strongly circularly polarized. In the resonance condition, the microwave frequency of the emission equals the cyclotron frequency  ($f_{B} = 2.80 \times 10^{6} B [\rm{Hz}]$, $B$ is the magnetic field strength in Gauss) or its harmonics \citep[2 or 3 times:][]{1992SoPh..141..303B}. When a resonant layer appears above the radio photosphere, it emits microwave radiation, whose brightness depends on the resonance-induced optical thickness, physical temperature at the resonance layer, and angle between the line of sight and magnetic field direction . Because of the magnetic field strength in the solar chromosphere and corona, the resonance layers above the radio photosphere appear only around the sunspots. Therefore, the source size of the gyroresonance emission is similar to the area of the sunspots and smaller than the Sun. \cite{2015ApJ...808...29S}  observed a full Sun at 2.8 GHz with the Karl G. Jansky Very Large Array (JVLA) on December 9, 2011 , corresponding to the early rising phase of Solar Cycle 24 and reported that the contribution of the gyroresonance emission to the total flux was only 8 \%. The gyroresonance emission spectrum depends on the magnetic fields above the radio photosphere and height structure of the temperature; the peak of the spectrum between 3 and 6 GHz in the Sun \citep{1994SoPh..152..167S}.

Based on the properties of the solar microwaves emitted from thermal plasma, we can estimate the average brightness temperature of microwave of a solar-type star during nonflaring periods. This is similar to the physical temperature in the radio photosphere at the observation frequency. This temperature is in the range of some thousands -- a hundred thousand K.  However, the flux density of stars calculated from the predicted brightness temperature is weak and insufficient for detection by using previous radio telescopes.

The state-of-art radio interferometers, the JVLA and the Atacama Large Millimeter/submillimeter Array, changed the situation. Some researchers have reported the detection of microwaves and millimeter/submillimeter-wave emissions from thermal plasma in the chromosphere and corona of main-sequence stars using these instruments  \citep{2014ApJ...788..112V, 2015A&A...573L...4L, 2016A&A...594A.109L, 2018ApJ...857..133B, 2018MNRAS.481..217T, 2020ApJ...904..138S}. Unfortunately, such observations are limited to the nearest stars even when interferometers are used. However, as the sensitivity of the planned radio interferometers to be realized in the next decade, such as the Square Kilometer Array \cite[SKA:][]{2009IEEEP..97.1482D} and the next-generation Very Large Array \citep[ngVLA:][]{2018SPIE10700E..1OS}, will be improved by tens or more compared with those of current instruments, we can predict that stellar radio astronomy, especially studies of the stellar chromosphere and transition region, will be a more activate research topic and the magnetic activities of many main-sequence stars will be investigated with microwaves.

The relationships between the total microwave flux and other activity indices of the Sun have been investigated mainly by using the F10.7 index because it has been used as one of the de facto standards, similar to sunspot numbers, for investigating long-term solar variations \citep[e.g.,][]{2017SoPh..292...73T, 2022ApJ...927..179T}. However, studying the relationship of microwaves with multiple frequencies is rare \citep[e.g.,][]{2018SpWea..16..434Z} because microwave data observed continuously with multiple frequencies over solar cycles are rare, especially after solar observations with satellites have become common. Moreover, no investigation has been conducted on the long-term variation in the circular-polarization degrees in the microwave range, which indicates the variation in solar magnetism. Therefore, preparations for new stellar radio astronomy in the next decade are not yet sufficient, and multifrequency templates created based on solar microwave observations are essential.

Japan has been monitoring the total microwave fluxes of the Sun and their circular polarization since the 1950s \citep{1984eyra.book..335T, 2014JAHH...17....2N}. Monitoring observations have continued until the present with multiple observation frequencies by using Nobeyama Radio Polarimeters  \citep[NoRP:][]{Tanaka1957, 1985PASJ...37..163N,2023GSDJ...10..114S}. The observing data of the NoRP have been primarily used for solar flare studies, but they can also be used to investigate long-term variations \citep[e.g.,][]{1994SoPh..152..167S,2017ApJ...848...62S}. In this study, we compared solar microwave data at multiple frequencies with other indices of solar activity and revealed their relationships. These relationships would help investigate stellar activity, at least for solar-type stars.

In the next section, we describe the datasets used in this study and the additional calibrations. We then present the relationships of the solar microwave fluxes and their polarization degrees with other solar indices. Finally, we provide conclusions regarding the obtained solar relationships and discuss the application of our results to future stellar microwave data.

\section{Long-term indices of solar activity} \label{sec:dataset}

In this study, we investigated solar activity using microwave data obtained with the NoRP and various solar indices. All of them have been released to the public, but we performed additional calibrations on some datasets. This section describes the datasets and additional calibrations. 

\subsection{Total microwaves fluxes and their circular polarization degrees observed with the NoRP} \label{sec:NoRP}

The NoRP is a cluster of solar radio telescopes for measuring the total fluxes and their circular-polarization degrees at 1, 2, 3.75, 9.4, 17, 35, and 80 GHz. Continuous monitoring observations have been conducted for over seven decades for 3.75 GHz \citep{Tanaka1953}. Although the telescopes are currently placed on the Nobeyama campus of the National Astronomical Observatory of Japan (NAOJ), monitoring observations with some frequencies did not start in Nobeyama. Because  \cite{2023GSDJ...10..114S} described the history of the NoRP, we do not provide such details in this paper and explain only the NoRP observation data used in this study. 

\begin{figure}
    \centering
    \includegraphics[scale=0.5]{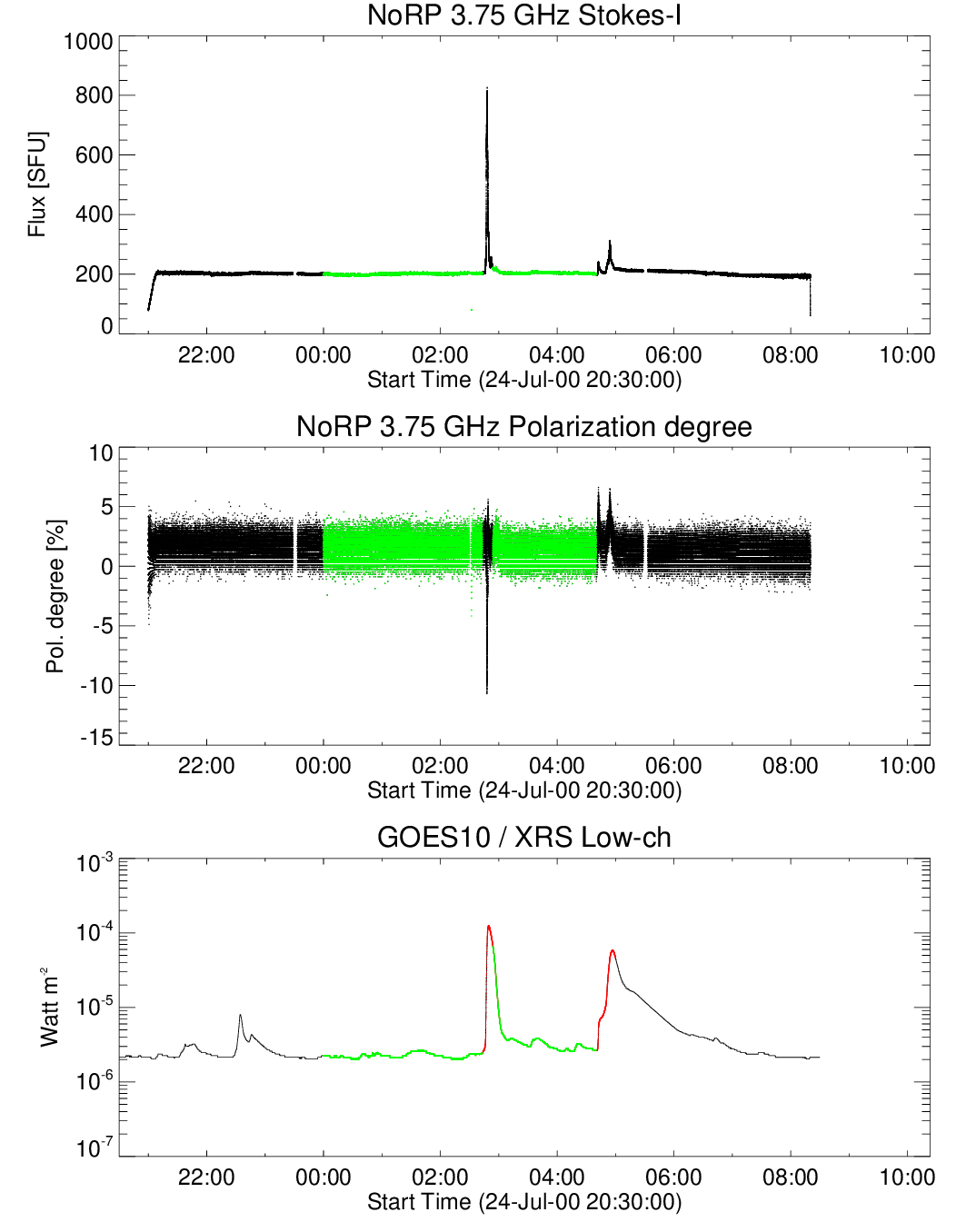}
    \caption{Time profiles on 25 July 2000. From top to bottom: 3.75 GHz flux, polarization degree of 3.75 GHz, and GOES X-ray flux. The red lines in the X-ray flux indicate the duration of the flares identified by the "pr\_gev" procedure, and the green dots in the panels show the data to estimate the average values.} 
    \label{fig:exDay}
\end{figure}

The Solar Data Archive System \citep[SDAS:][]{2018ASSL..449..247S}\footnote{https://hinode.nao.ac.jp/SDAS/} operated by the Astronomy Data Center and Solar Science Observatory of the NAOJ has provided tables of the daily solar fluxes at 1, 2, 3.75, 9.4, and 17 GHz, which were estimated by employing manually selected data. These tables are very useful for investigating solar cycles with microwaves because the fluxes are derived from data during the non-flaring period \citep[e.g.,][]{2017ApJ...848...62S}. However, these tables do not include circular-polarization information. Thus, the degree of polarization should be derived. We used the NoRP FITS database stored in the SDAS \citep{2023GSDJ...10..114S}\footnote{https://solar.nro.nao.ac.jp/norp/html/fits\_png\_ql/}. The database includes NoRP observational data starting from June 1, 1994. The FITS file of the database stores the 100-ms cadence Stokes-I and V fluxes observed in one day, and the fluxes are calibrated using the standard calibration method of the NoRP (Tanaka et al. 1973; Nakajima et al. 1985). Although the FITS file stores data with all observation frequencies of the NoRP, we used only 1, 2, 3.75, and 9.4 GHz data in this study because the effect of weather conditions cannot be neglected for data with a higher frequency ($>$10 GHz). 

Even when lower-frequency data are used, they include various spurious effects caused by weather conditions, ground effects, and instrumental issues. Moreover, we need to avoid solar phenomena related to nonthermal electrons. To remove these effects, we used data that satisfied the following conditions to obtain the average values for a day  (Figure \ref{fig:exDay}).

\begin{enumerate}
\item When the solar flux was not recorded in the table of daily solar fluxes, the average value of the day was not estimated. Lack of the daily flux derived manually means that the instrument did not work well or that the weather conditions were extremely bad; for example, snow was present on the surface of the antenna. 
\item We used data observed between 00:00UT (9:00JST) and 5:00UT (14:00JST). When the elevation of the Sun was low, the microwaves reflected by the ground interfered with the microwaves coming directly from the Sun. 
\item Because we investigated microwave emissions from thermal plasma, we removed the data obtained over the C-class flares listed in the GOES event list. The event list was obtained using the “pr\_gev” procedure in Solar Software \citep[SSW:][]{1998SoPh..182..497F}. 
 \end{enumerate}

\begin{figure*}
    \centering
    \includegraphics{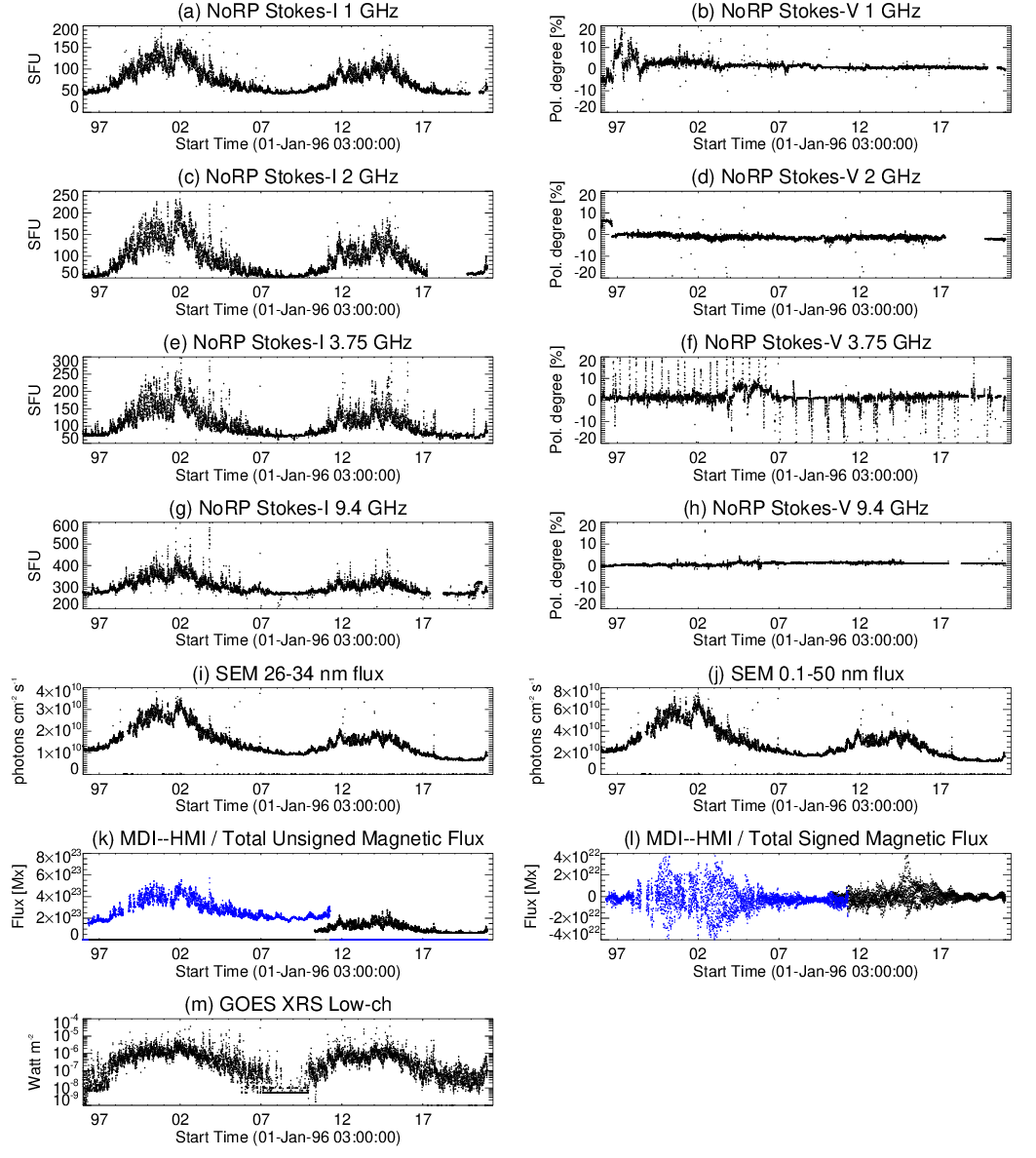}
    \caption{Time profiles of the original data}
    \label{fig:org}
\end{figure*}

Even when we applied Condition \#3 to the 1 GHz data, we could not remove the effect of metric bursts and maser emissions because the periods of these emissions were delayed from the flare periods determined from the GOES X-ray data. Hence, we do not use 1 GHz data in which the flux exceeds 300 SFU\footnote{Solar Flux Unit. 1 SFU = 10$^{4}$ Jy = 10$^{-22}$ W m$^{-2}$ Hz$^{-1}$} because such a high value would indicate these phenomena. 

Plots (a) -- (h) in Figure \ref{fig:org} show the long-term solar variation with microwaves from January 1, 1995, to December 31, 2020. Each dot in the plot indicates the average solar flux and averaged circular-polarization degree of a day, estimated from data that satisfy all the aforementioned conditions.

\begin{figure*}
    \centering
    \includegraphics{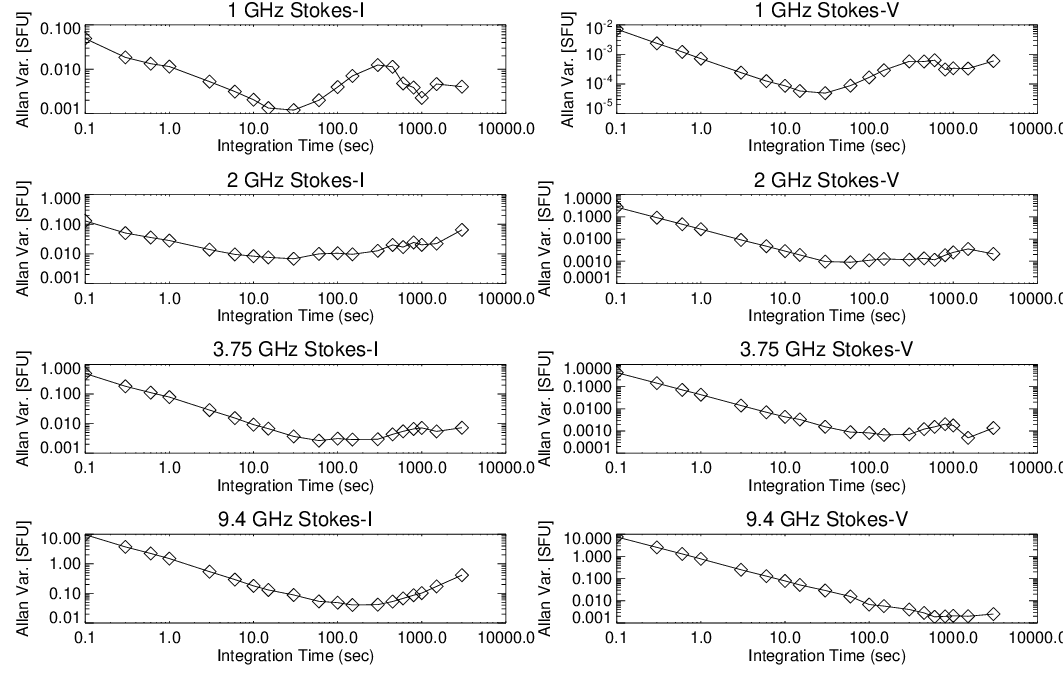}
    \caption{Allan variations of the microwave fluxes and circular polarization degrees}
    \label{fig:allan}
\end{figure*}

To date, the detection limit of the circular-polarization signal observed with the NoRP has not been seriously considered because the circular-polarization degree of a radio-loud solar flare, the primary scientific target of the NoRP, usually exceeds a few tens of percent. However, to study the polarization at non-flaring times, we should first investigate the polarization accuracy of NoRP. For this purpose, we checked the Allan variances of the NoRP Stokes-I and V data on the quietest day of the solar minimum. Figure \ref{fig:allan} shows the Allan variances of the NoRP data. The representative value the Allan variances of the Stokes-I and V are about 0.1 SFU and 0.01 SFU at the integration time.

Given the stability evaluated from Allan variances, the statistical error in the circular-polarization degrees should be less than 1\%. However, we observe a significant variance in the plots of the polarization degrees in Figure \ref{fig:org}. Significantly, the large scatter in the 3.75 GHz appears frequently. This can be attributed to two reasons. The first is interference from geosynchronous satellites. A frequency range of 3,400 -- 4,200 MHz was used for the satellite communication. The line of sight of the Sun passes through the geosynchronous satellite belt near the spring and autumn equinoxes. Owing to polarized emissions from geosynchronous satellites, the reliability of the Stokes V data in this period was significantly degraded. Another reason is also related to the geosynchronous satellites. Since 2006, the observation frequency has been regularly changed to avoid radio interference from geosynchronous satellites \citep{2023GSDJ...10..114S}. Because the polarization splitter was optimized for a fixed frequency, the frequency change generated pseudo-polarization signals. The Stokes I fluxes at different observation frequencies were calibrated using the microwave spectrum. However, such treatments cannot be applied to Stokes V fluxes. Because of these issues, we did not use the 3.75 GHz data from September to May.

\begin{figure*}
    \centering
    \includegraphics{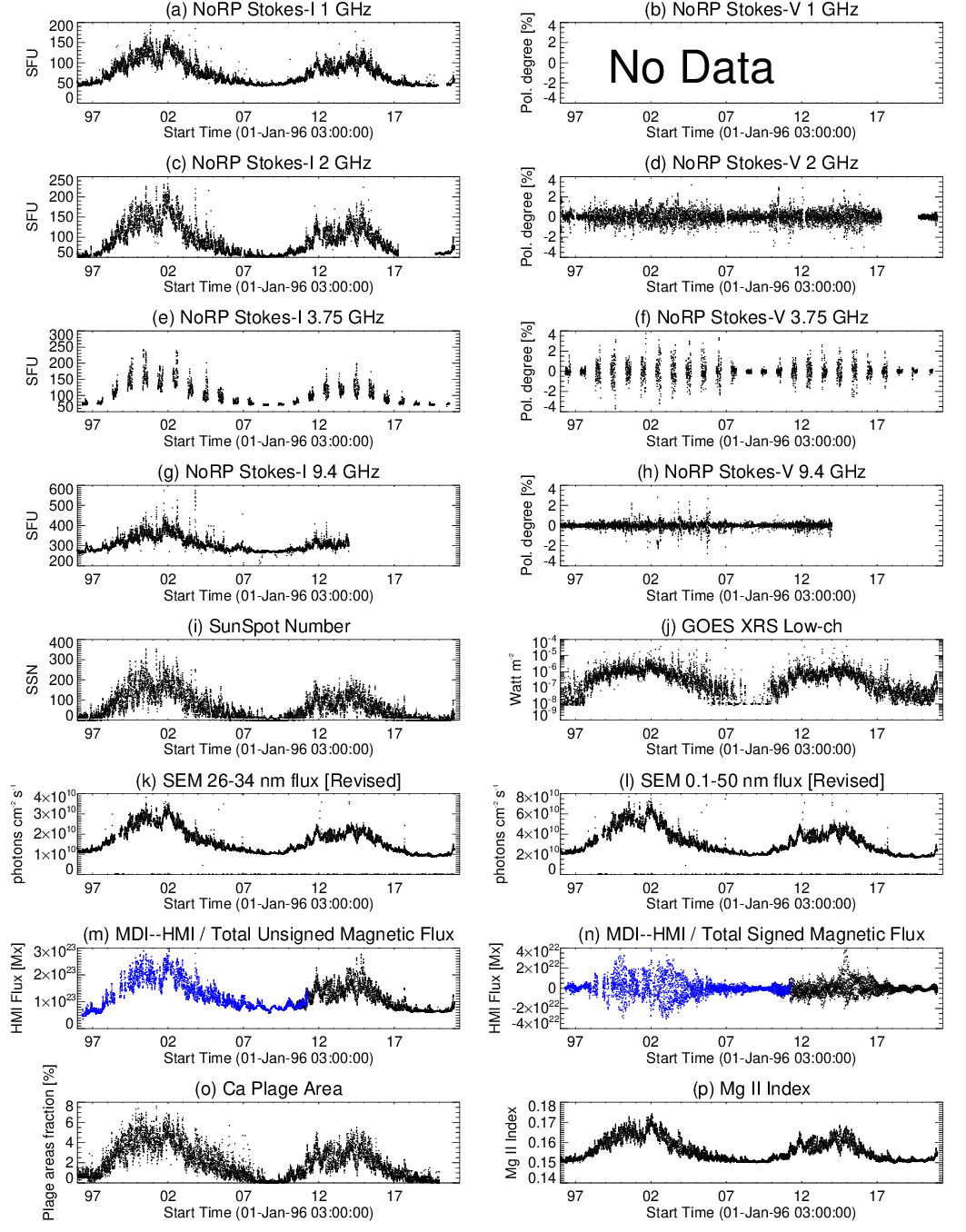}
    \caption{Time profiles of the calibrated data}
    \label{fig:cal}
\end{figure*}

We can see significant variations of polarization degrees in all frequencies even when we removed the 3.75 GHz data from September to May. As shown in the Stokes-V time profiles in Figure  \ref{fig:org}, the values did not align around zero, even when there was no active region. This indicates that systematic errors were present. To calibrate the error, we assumed that the monthly average of the circular-polarization degree was zero, calculated the actual monthly averages of the polarization degrees as the offset, and calibrated the data using the offset values. The calibration results are shown in plots (d), (f), and (h) in Figure  \ref{fig:cal}.

The polarization degree of 1 GHz scattered even after additional calibration, and the variations were not related to the solar cycles. We checked the logbook of the NoRP operation and found that polarization tuning was frequently performed after moving the 1 GHz antenna from Toyokawa. Therefore, we conclude that the accuracy and stability of the 1 GHz polarization data are insufficient for our study and give up using the polarization data at 1 GHz. 

\cite{2023GSDJ...10..114S} reported that the 9.4 GHz antenna had lost the function of measuring polarization from September 18, 2014, to June 28, 2021. Therefore, we do not use the 9.4 GHz data of the period. 

To estimate the uncertainties in the calibrated NoRP data used in this study, we estimated the standard deviation of the Stokes I fluxes and circular-polarization degrees during the deepest solar minimum. We selected June -- August 2008 as the deepest solar minimum because it had the lowest activity period of cycle 23 in the microwaves \citep{2017ApJ...848...62S}. The standard deviations for the Stokes-I fluxes during the period are 1.32 SFU (2.9 \% of the fluxes) for 1 GHz, 0.51 SFU (1.0 \%) for 2 GHz, 0.48 SFU (0.7 \%) for 3.75 GHz, and 1.67 SFU (0.6 \%) for 9.4 GHz. The standard deviations of these circular-polarization degrees were 0.25 \% for 2 GHz, 0.1 \% for 3.75 GHz, and 0.1 \% for 9.4 GHz. The uncertainties of the calibrated NoRP data would not significantly exceed these values, except for the period of instrumental problems. 

\subsection{EUV fluxes observed with the Solar EUV Monitor (SEM)}\label{sec:sem}

The SEM is a part of the Charge, Element, and Isotope Analysis System \citep[CELIAS:][]{1995SoPh..162..441H}  aboard the SOlar and Heliospheric Observatory \citep[SOHO:][]{1995SoPh..162....1D}, and continuously measures the EUV flux in a broad band of 0.1 -- 50 nm and a narrow band around the 30.4 nm (He II). The University of Southern California\footnote{The archive and providing website of SEM was moved to the website of the Laboratory for Atmosphere and Space Physics (LASP), University of Columbia.} provided calibrated SEM data, and we used the 15-s cadence data (Version 3b) for our study. We estimated the average EUV flux per day from the data for the same period used to calculate the average value of the microwaves. The long-term variations in the averaged EUV fluxes are shown in plots (i) and (j) of Figure \ref{fig:org}. 

When we see the periods of low solar activity in the plots (i) and (j) of Figure \ref{fig:org}, the EUV fluxes at the solar minima decreased significantly. However, a decreasing trend does not indicate actual solar variation. \cite{2017JGRA..122.3420H} compared SEM data and EUV fluxes obtained from the Solar EUV Experiment (SEE) aboard the TIMED satellite and revealed that the sensitivity of the SEM was degrading. They suggested a temporal sensor drift coefficient of approximately -2 \% per year.  \cite{2014SoPh..289.2907W} revealed that the result of \cite{2017JGRA..122.3420H} are consistent with the difference between the SEM EUV fluxes and the EUV fluxes derived  by   from the  data obtained from an Extreme Ultraviolet Variability Experiment  \citep[EVE:][]{2012SoPh..275..115W} aboard the Solar Dynamic Observatory \citep[SDO:][]{2012SoPh..275....3P}. Hence, the formula described by \cite{2017JGRA..122.3420H}  was used to revise the SEM EUV flux. The revised data are shown in plots (k) and (l) of Figure \ref{fig:cal}. Even when we revised the data, the EUV flux at the solar minimum between cycles 23 and 24 was approximately 6 -- 8\% smaller than at the solar minimum between cycles 22 and 23. We do not discuss the solar minimum between cycles 24 and 25, which is discussed later in this paper.

\subsection{Total magnetic fluxes obtained with the Michelson Doppler Imager (MDI) and the Helioseismic and Magnetic Imager (HMI)}\label{sec:MAG}

The total magnetic flux is essential for investigating long-term solar activity because most solar activity originates from magnetic fields. We used two datasets to calculate the total magnetic flux. One is the magnetogram observed with the MDI \citep{1995SoPh..162..129S} aboard the SOHO. The MDI team provides FITS files of the calibrated magnetograms (Levels 1.8.2) at their FTP site\footnote{ftp://soi-ftp.stanford.edu/pub/magnetograms/}. We used FITS files that stored the magnetograms observed with a 5-min integration time at approximately 3:00 UT, which was noon in Japan. The MDI dataset covers April 15, 1996, to April 15, 2011. The other dataset comprises the synoptic magnetograms observed with a HMI \citep{2012SoPh..275..207S}aboard the SDO. We obtained the HMI FITS file of the synoptic magnetogram observed at approximately 3:00UT from the Joint Science Operation Center (JSOC)\footnote{http://jsoc.stanford.edu/data/hmi/fits/}. The HMI data used in this study cover the period from May 1, 2010, to December 31, 2020. 

After obtaining the FITS file from the data archive site, we first removed the pixels where the distance from the disk center is over 0.95 solar radius were removed, and the magnetic flux density was less than the noise level \citep[15 Gauss for MDI: SOI-Technical Note \#01-144\footnote{http://soi.stanford.edu/technotes/01.144/TN01-144.pdf}, 5 Gauss for HMI:][]{2020ApJ...902...36T}. We then calculated the physical surface area of each pixel from the distance between the pixel and the disk center and estimated the magnetic flux in each pixel. In the calculation, we considered seasonal variations in the distance from the Sun to each telescope. Subsequently, we summed the magnetic fluxes of all the magnetized pixels to obtain the total signed magnetic flux. To obtain the total unsigned magnetic flux, we first obtained the absolute value of the magnetic flux of each pixel and then summed the values of all the magnetized pixels. Plots (k) and (l) in Figure  \ref{fig:org} show the long-term variations in these values. In the plots, the blue dots indicate MDI data and the black dots indicate HMI data. This study used this color code for all magnetic flux plots. As we did not convert them to radial fluxes, the values were the total unsigned/signed line-of-sight magnetic fluxes. 

As shown in plots (k) and (l) in Figure \ref{fig:org}, gaps exist between the values obtained from MDI and HMI. This is natural because there are differences not only in instruments, but also in the observing lines. However, these datasets are not suitable for studying long-term variations. Since the two datasets overlapped from May 1, 2010, to April 15, 2011, we have revised the data using the overlapping period. First, we created scatter plots of the total unsigned/signed magnetic fluxes during the overlapping period and fitted the scatter plots with a linear function. We then revised the total unsigned/signed magnetic fluxes obtained from the MDI magnetograms using the fitting parameters. The revised values are shown in plots (m) and (n) of Figure \ref{fig:cal},. We revise only the MDI data. Therefore, we refer to the revised fluxes as ``HMI Flux'' in this paper. 

\subsection{X-ray flux obtained with the X-Ray Sensor (XRS)}\label{sec:XRS}

The solar X-ray flux observed with the low channel of the XRS aboard GOES satellites is usually used to evaluate solar activity, especially for solar flares, and is the de facto standard in the field. We used the GOES SSW package \citep{1998SoPh..182..497F} to obtain the GOES X-ray data. To calculate the average values of the X-ray flux per day, we used the same period to estimate the average of the microwave fluxes to calculate the average value of the X-ray fluxes. This is the same as the treatment for the EUV fluxes. Plot (m) in Figure \ref{fig:org} shows the time profile of the values.

The X-ray fluxes around the solar minimum of cycle 24 (2007 -- 2010) may have been lower than the detection limit of the instrument. Hence, we used only the X-ray fluxes larger than  8 $\times$ 10$^{-9}$ Watt m$^{-2}$. Plot (j) of Figure \ref{fig:cal} shows the data used in this study. 

\subsection{Other indices of solar activity}\label{sec:others}

We also used sunspot numbers, the Ca II plage area fraction, and the Mg II index in this study. Their time profiles are shown in plots (i), (o), and (p) of Figure \ref{fig:cal}. Daily sunspot numbers were obtained from the WDC-SILSO, Royal Observatory of Belgium\footnote{https://wwwbis.sidc.be/silso/home/}.

\cite{2020A&A...639A..88C} produced a plage areas time series from the Ca K II images observed worldwide since 1892 and released the Strasbourg astronomical Data Center (CDS)\footnote{https://cdsarc.cds.unistra.fr/viz-bin/cat/J/A+A/639/A88}. The value in the dataset is the rate of total plage areas in the solar disk. Therefore, we called the value ``Ca plage fraction" in this paper.  

Mg II indices were provided at the Interactive Solar Irradiance Data Center\footnote{https://doi.org/10.25980/L27Z-XD34}, operated by the Laboratory for Atmosphere and Space Physics, University of Columbia. Several Mg II indices are present in the data center. Among these indices, we used the Bremen composite Mg II index \citep{2014JSWSC...4A..04S} because the time-series data covered the entire period of our study. As per the explanation page of the Bremen composite\footnote{https://www.iup.uni-bremen.de/gome/gomemgii.html}, the NoRP 1 GHz data may have been used for the composite. We asked the contact person of the Bremen composite about the issue and confirmed that the NoRP data were not used to composite the recent solar cycles but were used only to restore the index in the 1980s.

\section{Relationships between the solar microwaves data and other solar indices}\label{sec:result}

We investigated the microwave data obtained using NoRP and the other indices of solar activity described in the previous section (Figure  \ref{fig:cal}). In this section, we first elucidate the relationships between the microwave fluxes and other indices and then compare the circular-polarization degrees of microwaves with the other indices strongly related to magnetic fields.

\subsection{Microwave fluxes and the other indices}\label{sec:fluxIndexes}

Microwave fluxes were compared with X-ray fluxes during non-flaring periods using the F10.7 index and X-ray data obtained with the spacecraft in the late 1960s \citep{1969JGR....74.4649W, 1970ApJ...161.1135G}. Then, \cite{1982JGR....87.6331D} investigated this relationship using the F10.7 index and X-ray flux obtained from the GOES 1 and GOES 2 satellites. First, the results of Donnelly (1982) were reproduced using data obtained from NoRP and XRS/GOES.

\begin{figure*}
    \centering
    \includegraphics[scale=0.9]{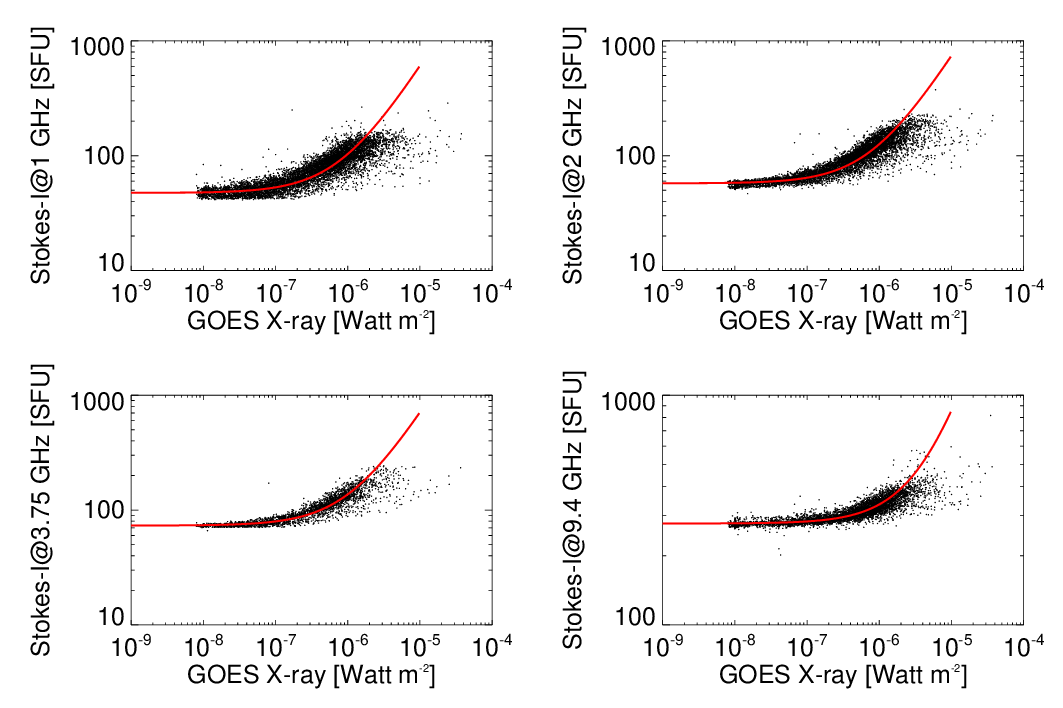}
    \caption{The relations between the solar microwave and X-ray fluxes}
    \label{fig:RadioXray}
    
    \includegraphics[scale=0.9]{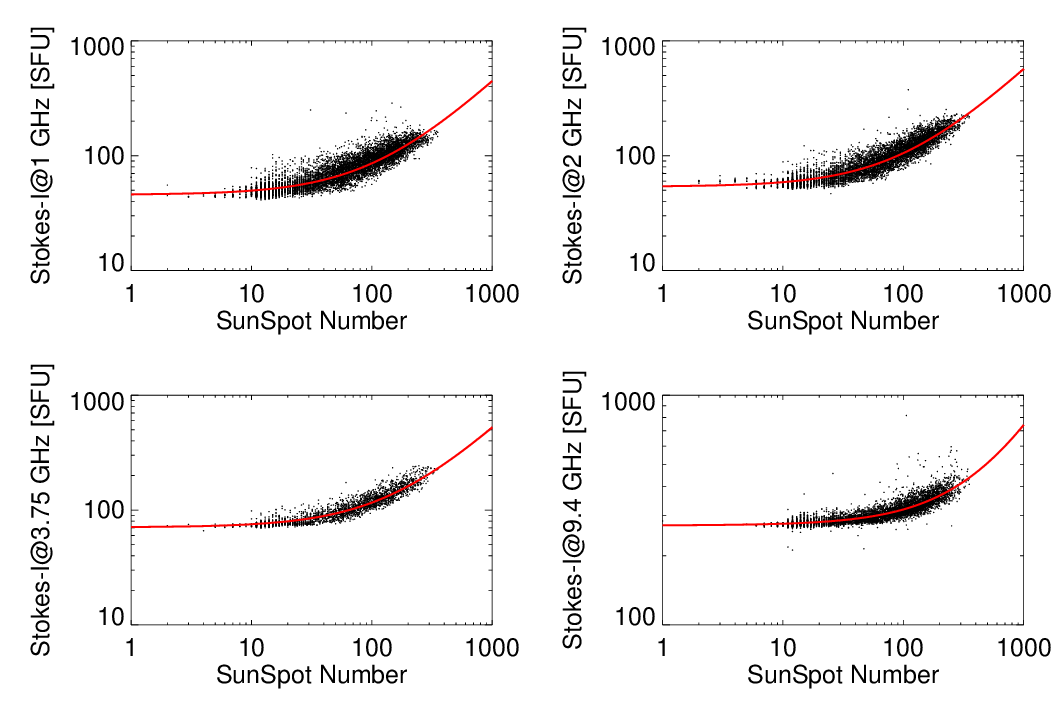}
    \caption{The relations between the solar microwave fluxes and sunspot number}
    \label{fig:RadioSunspot}
\end{figure*}

\begin{figure*}
    \centering
    \includegraphics[scale=0.9]{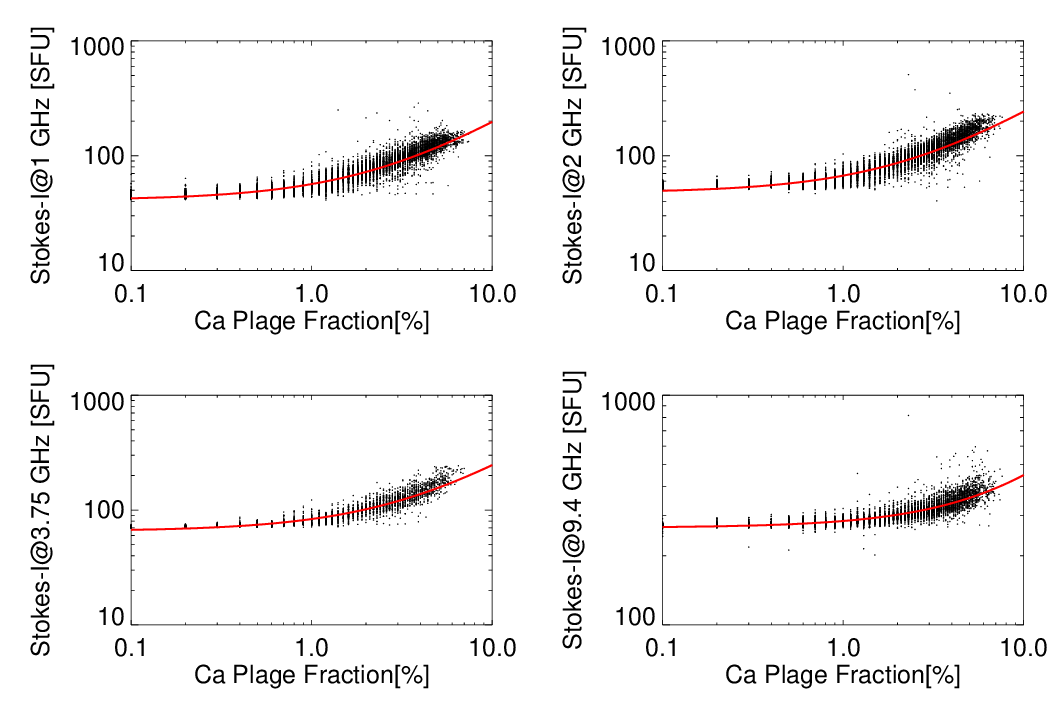}
    \caption{The relations between the solar microwave fluxes and Ca II Plage Fraction}
    \label{fig:RadioCaII}
    
    \includegraphics[scale=0.9]{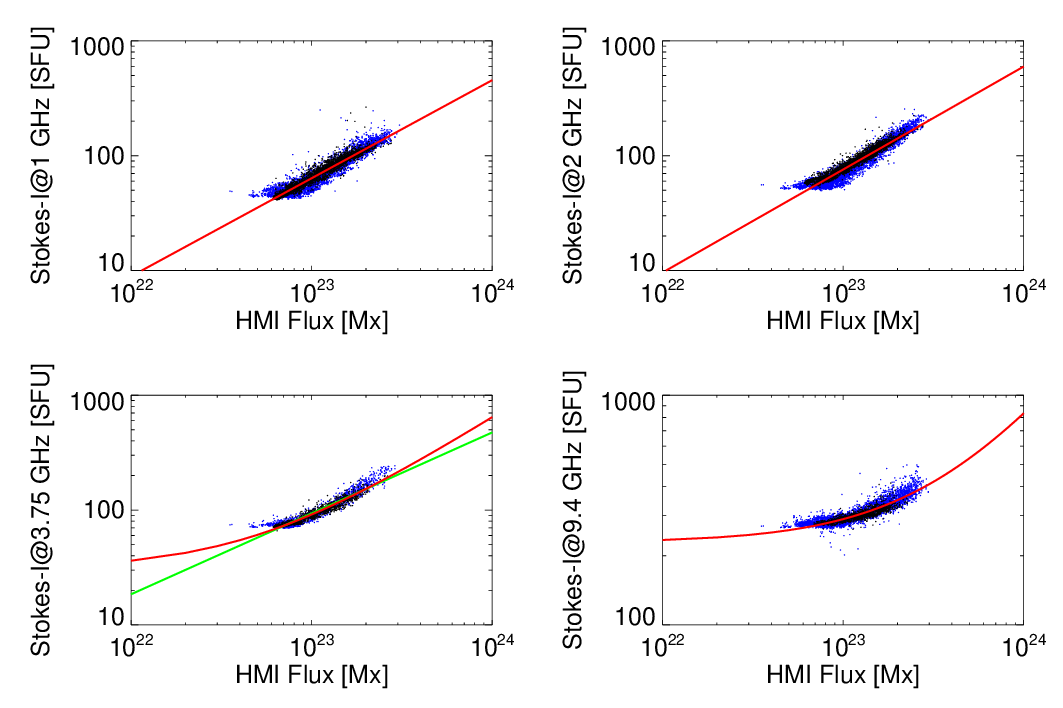}
    \caption{The relations between the solar microwave fluxes and total unsigned magnetic flux. The red and green lines in the lower-left panel (3.75 GHz) show the results of the fitting with a linear (red) and power-low (green) function respectively.}
    \label{fig:RadioMag}
\end{figure*}

\begin{figure*}
    \centering
    \includegraphics[scale=0.7,angle=90]{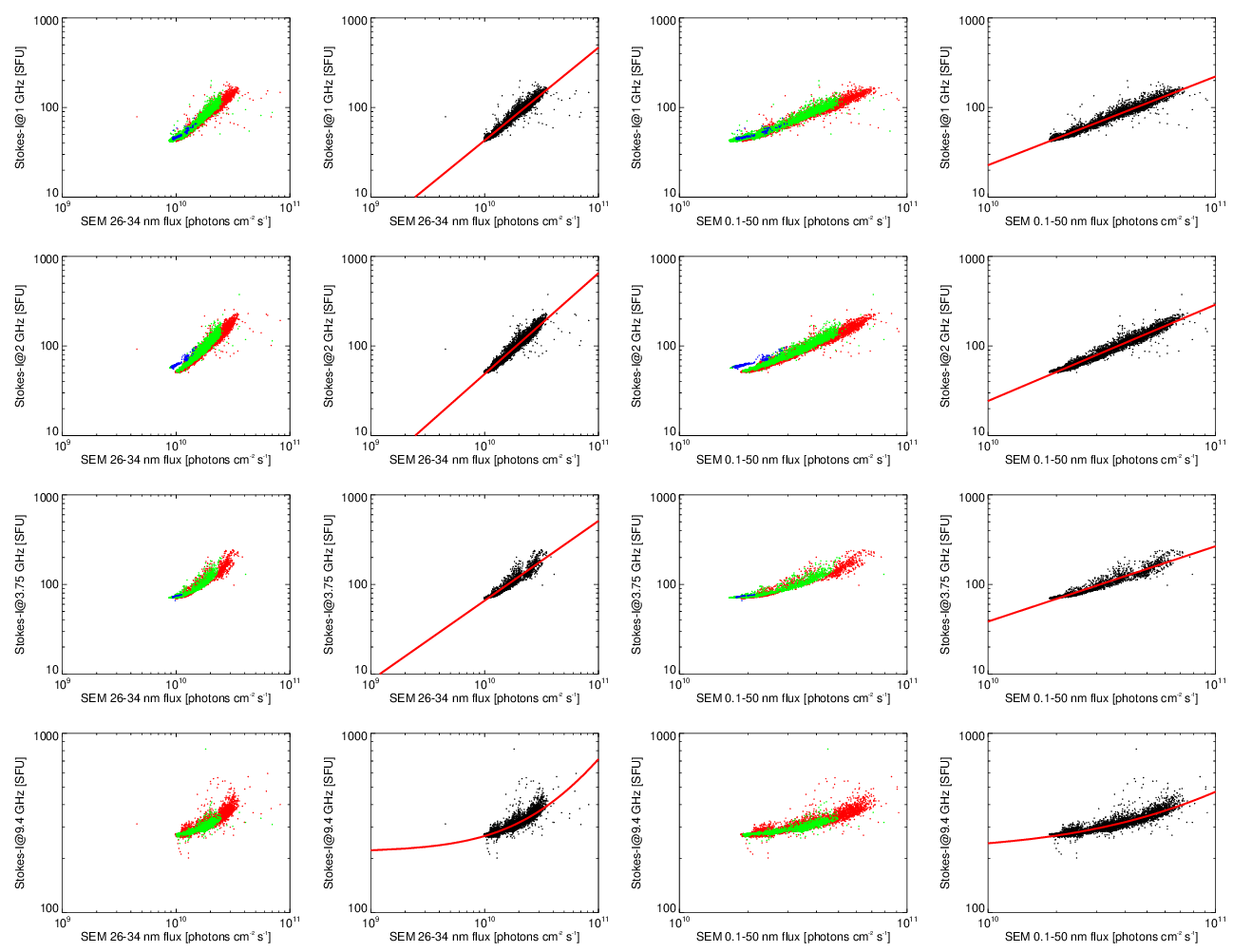}
    \caption{The relations between the solar microwave and EUV fluxes. The plots of 1st and 3rd rows are produced the EUV data obtained during the whole period of this study; Red dots: Cycle 23, Green dots: Cycle24, Blue dots: Cycle 25. The plots of 2nd and 4th plots are produce from the data obtained until the end of 2013.}
    \label{fig:RadioEUV}
\end{figure*}

\begin{table*}

\centering
\caption{Results of the fitting with a linear function ($ F_{\rm microwave} = a + b \times I_{\rm index}$). The errors are in 1-sigma uncertainty. We used the values presented in the last paragraph of Section \ref{sec:NoRP} as the measurement errors of the microwave fluxes in the error estimation.} 
\label{tab:fitResLin}

\begin{tabular}{c|cc|cc}
\hline\hline
&\multicolumn{2}{c}{1 GHz flux}&\multicolumn{2}{c}{2 GHz flux}\\
&a&b&a&b\\
\hline
X-ray flux & $47.40 \pm   0.03$ & $(5.61\pm0.01) \times 10^ {7}$ & $57.46 \pm     0.01$ & $(6.86 \pm 0.01) \times 10^{7}$ \\ 
Sunspot number & $45.64 \pm 0.02$ & $0.402 \pm 0.001$ & $53.68 \pm  0.01$ & $0.516 \pm   0.001$ \\
Ca plage & $41.01 \pm    0.03$ & $15.62 \pm    0.02$ & $47.58 \pm    0.01$ & $19.49 \pm    0.01$ \\
Mg II Index & $-738.8 \pm   0.8$ & $5200 \pm   5$ & $-918.5 \pm   0.4$ & $6444 \pm  2$\\
\hline\hline
&\multicolumn{2}{c}{3.75 GHz flux}&\multicolumn{2}{c}{9.4 GHz flux}\\
&a&b&a&b\\
\hline
X-ray flux & $73.24 \pm     0.02$ & $(6.38 \pm 0.06) \times 10^{7} $ &  $276.3 \pm     0.1$ & $(5.82 \pm0.01)\times10^{7}$\\ 
Sunspot number &  $70.67 \pm    0.02$ & $0.453 \pm   0.001$ & $271.0 \pm    0.1$ & $0.471 \pm   0.001$ \\
Ca plage &  $65.39 \pm    0.02$ & $18.09 \pm    0.01$ & $265.0 \pm    0.1$ & $18.41 \pm    0.02$ \\
Unsigned Mag. & $30.03 \pm     0.04 $ & $(6.144\pm0.004)\times10^{-22} $ & $228.1 \pm     0.1$ & $(6.060 \pm 0.005) \times 10^{-22}$\\
EUV 26-34nm Flux & - & - & $217.2 \pm     0.1$  & $(5.002 \pm 0.004) \times 10^{-9}$ \\

EUV 0.1-50nm Flux & - & - & $218.0 \pm     0.1$  & $(2.536 \pm 0.002)\times 10^{-9} $ \\
\hline

\end{tabular}

\caption{Results of the fitting with a power-law function ($F_{\rm microwave} = \alpha \times I_{\rm index}^{\beta}$).} 
\label{tab:fitResLog}

\begin{tabular}{c|cc|cc}
\hline \hline
 &\multicolumn{2}{c}{1 GHz flux}&\multicolumn{2}{c}{2 GHz flux}\\
& $\alpha$ & $\beta$ & $\alpha$ &  $\beta$  \\
\hline
Unsigned Mag.       & $1.45^{+0.27}_{-0.23}\times10^{-18}$ & $0.854 \pm    0.003$ & $1.76^{+0.34}_{-0.29}\times10^{-19}$ & $0.897 \pm    0.003$ \\

EUV 26-34nm Flux &$1.89^{+0.16}_{-0.15}\times10^{-9}$ & $1.036 \pm    0.004$ & $2.69^{+0.26}_{-0.23}\times10^{-10}$ & $1.126 \pm    0.004$  \\

EUV 0.1-50nm Flux & $2.88^{+0.21}_{-0.20}\times10^{-9}$ & $0.990 \pm    0.003$ & $4.30^{+0.34}_{-0.32}\times10^{-10}$ & $1.075 \pm    0.003$  \\
\hline\hline
&\multicolumn{2}{c}{3.75 GHz flux}&\multicolumn{2}{c}{9.4 GHz flux}\\
\hline
Unsigned Mag.       & $5.79^{+1.50}_{-1.19}\times10^{-15}$ & $0.705 \pm    0.004$ & - &  - \\
EUV 26-34nm Flux & $8.37^{+1.36}_{-1.17}\times10^{-8}$ & $0.890 \pm    0.006$ & -  & - \\
EUV 0.1-50nm Flux & $1.51^{+0.22}_{-0.19}\times10^{-7}$ & $0.841 \pm    0.006$ & -  & - \\
Mg II Index & $1.76^{+0.16}_{-0.15}\times10^{9}$ & $8.990 \pm    0.048$ & $8.76^{+0.31}_{-0.30}\times10^{4}$ & $3.053 \pm    0.019$ \\
\hline
\end{tabular}

\end{table*}

Figure  \ref{fig:RadioXray} shows the scatter plots between the total microwave fluxes at 1, 2, 3.75, and 9.4 GHz and the X-ray fluxes. \cite{1982JGR....87.6331D} presented similar plots, but the lowest level of X-ray fluxes in his plot is 3 $\times$ 10$^{-8}$ Watt m$^{-2}$ (GOES A3 level), and the data points lower than 1 $\times$ 10$^{-7}$ Watt m$^{-2}$ (GOES B1 level) are rare. Hence, the relationship between the microwave and X-ray fluxes near the solar minima is unclear from the plot. Figure \ref{fig:RadioXray} includes the data of three solar minima and reveals that the variations in the microwave fluxes are negligible even when the X-ray flux changes by over one order, from  8 $\times$ 10$^{-9}$ Watt m$^{-2}$ to 1 $\times$ 10$^{-7}$ Watt m$^{-2}$.. When the X-ray flux is over 1 $\times$ 10$^{-7}$ Watt m$^{-2}$, the relation shows monotonically increasing, as  \cite{1982JGR....87.6331D} presented. The relationship can be fitted well with a linear function ,  ($F_{\rm microwave} = a + b \times I_{\rm index}$)and the results are shown as red lines in these plots and summarized in Table \ref{tab:fitResLin}.

This linear relationship also appears in relation to other indices. Figure \ref{fig:RadioSunspot} and Figure \ref{fig:RadioCaII} show scatter plots between the microwave fluxes and sunspot numbers, and between the microwave fluxes and the Ca plage fraction, respectively. The fitting results are shown as red lines in these plots and summarized in Table \ref{tab:fitResLin}. There relationships are the similar to the relationship between the microwave fluxes and  X-ray flux. 

Some relationships can be fitted well with a power-law function rather than a linear function. An impressive example is the relationship with the total unsigned magnetic flux (Figure \ref{fig:RadioMag}). The fitting results with a power-low function  ($F_{\rm microwave} = \alpha \times I_{\rm index}^{\beta}$)  are indicated with the red and green lines and index summarized in Table  \ref{tab:fitResLog}. As shown in Figure \ref{fig:RadioMag}, the relationship changes with the observed microwave frequency. The scatter plot with the 3.75 GHz flux can be fitted by using a power-law function, but the residual from the fitting result is relatively larger than that with a linear function. The 9.4 GHz flux plots can be fitted well with a linear function rather than a power-law function.

Such trends are evident in the relationship between the microwave fluxes and EUV fluxes. Before discussing the relationship with EUV fluxes, we note the revision of the EUV data after 2014. As mentioned in the previous section, the EUV fluxes used in this study were revised using the equations and parameters proposed by \cite{2017JGRA..122.3420H}. Moreover, the revised values in 2013 were consistent with the EVE data \citep{2014SoPh..289.2907W}. This means that the revised data until 2013 were verified, but the data after 2014 (after the middle of the rising phase of Solar Cycle 24) were not verified using other instruments. In Figure 4, the EUV fluxes at the minimum of Solar Cycle 25 (blue dots) are significantly lower than those in cycles 23 and 24(red and green dots), although the other indices do not show such differences. Moreover, the EUV fluxes around the solar maximum of Solar Cycle 24 on the scatter plots with 1 and 2 GHz fluxes are saturated (green dots in the 1st and 3rd column of Figure  \ref{fig:RadioEUV}). These trends suggest that the revisions did not work well after 2014.  \cite{2022SoPh..297...43W}  reported no significant changes in the EUV fluxes between the solar minima of cycles 24 and 25. Considering these results, we used only the EUV data obtained at the end of 2013 for our examination.

As previously mentioned, the relationship between the microwave and EUV fluxes was similar to that between the microwave and total unsigned magnetic fluxes. The scatter plots (the 2nd and 4th column in Figure  \ref{fig:RadioEUV}) can be fitted well with a power-law function, except for 9.4 GHz data. The fitting results with a linear function and a power-law function are summarized in Tables  \ref{tab:fitResLin} and \ref{tab:fitResLog}, respectively.

\begin{figure*}
   \centering
    \includegraphics[scale=0.9]{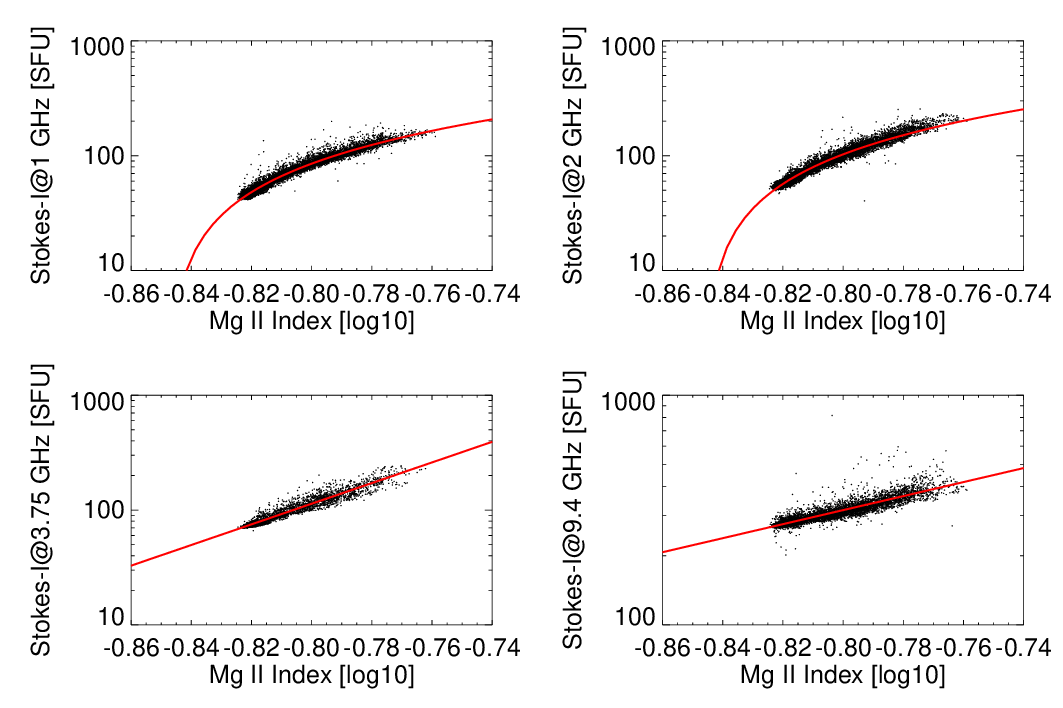}
    \caption{The relations between the solar microwave fluxes and Mg II index}
    \label{fig:RadioMgII}
\end{figure*}

The relationship between the microwave flux and Mg II index is shown in Figure \ref{fig:RadioMgII}. The plots show opposite properties from the relation between the EUV and unsigned magnetic fluxes, and the relations between the 1 and 2 GHz fluxes and the Mg II index fit well with a linear function. Conversely, the relations of the Mg II index with the 3.75 and 9.4 GHz show the power-law functions.

\subsection{Circular-polarization degrees of microwaves and the indices}

\begin{figure}
    \centering
    \includegraphics[scale=0.9]{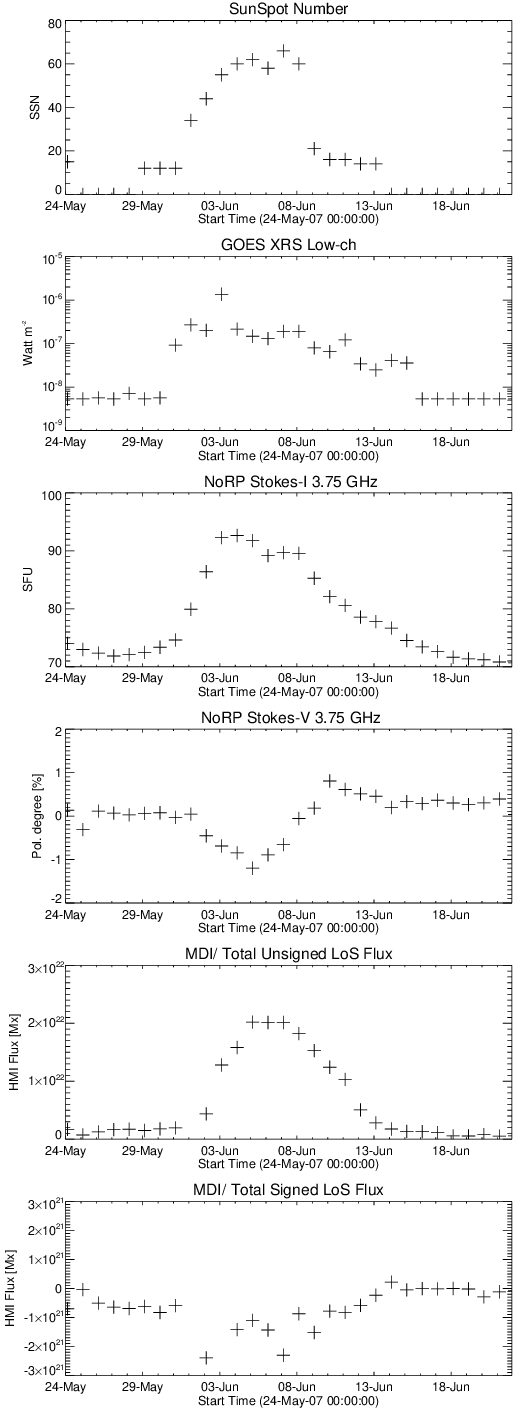}
    \caption{The time profiles of the microwave and total magnetic fluxes ($>$250 Gauss) during NOAA10960 passing the solar disk.}
    \label{fig:noaa10960}
\end{figure}

\begin{figure}
    \centering
    \includegraphics[scale=1.0]{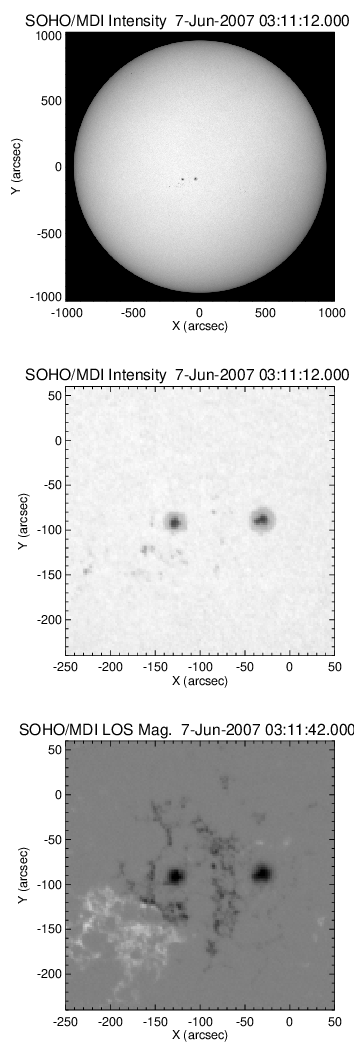}
    \caption{The Sun observed with MDI/SOHO on 7 June 2007. From Top, Full Sun Intensity, NOAA10960 Intensity, and NOAA10960 LOS magnetogram.}
    \label{fig:960sun}
\end{figure}

Before discussing the relationship between the long-term indices of solar activity and the circular-polarization degree of microwaves, we introduce the behavior of the polarization degree in an isolated active region passing from east to west on the solar disk, which is the simplest case. Figure \ref{fig:noaa10960} shows the time profiles of the solar indices when an active region “NOAA10960” passed from the east limb to the west limb (May 24 -- June 22, 2007; the date of meridian passing is June 7). During this period, no other active regions were present except for NOAA10960. The total magnetic fluxes in the figure (4th and 5th row in Figure \ref{fig:noaa10960}) are calculated from the pixels where the magnetic field strength is over 250 Gauss to emphasize the fluxes of the active region. The active region had two large sunspots with negative polarities (Figure \ref{fig:960sun}). Two distinct sunspots with the same polarity in an active region are not very common, but we can imagine that the situation is the same as when a large sunspot passes on a solar disk from our point of view because we discuss only the total flux and its polarization degrees. The magnetogram shows a region with a positive polarity on the southeast side of the sunspots. In the section, we refer to the region as ``positive region.''

The resonance layers of the gyroresonance emissions were created above the sunspots and positive region. The area of the resonance layer depends on the strength of the photospheric magnetic field; a strong magnetic field region creates wider layers above it. Hence, the microwave flux emitted above the sunspots is larger than that above the positive region.

The circular polarities of the microwaves from the sunspots and positive regions are opposite. The polarization degree obtained using NoRP indicates the summed polarization degrees of these signals. The magnetic field strengths of the preceding spots are usually stronger than those of the following spots or regions. Hence, the total signed magnetic flux became negative when NOAA 10960 appeared near the eastern limb on 30 May. Similarly, the degree of circular polarization became negative because the emission from the resonance layers above the sunspots was more prominent than that above the positive region. The microwave flux increased owing to the gyroresonance emission from above the sunspots, and the X-ray flux and sunspot number also increased.

As shown in the rising phase of NOAA10960, the circular-polarization degrees of microwaves and the other indices showed similar behaviors until June 7. The behavior of the circular-polarization degree becomes complex after the AR passes through the meridian. When the active region approaches the west limb, the circular polarization becomes positive and then decreases to approximately zero.

As  \cite{1997SoPh..174...31W}  described, changing the viewing angle can dramatically affect polarization. The degree of polarization increases when a sunspot is observed just above it. However, when we observe this from the side, the degree of polarization becomes significantly lower. After the sunspots passed the meridian, we observed the positive region just above it and the sunspots from the side. Circular polarization might change to a positive value, mainly because of the viewing-angle effect. As shown in the simple case, the relationship between the circular-polarization degree and the solar indices is complex.

\begin{figure*}[t]
    \centering
    \includegraphics[scale=0.9]{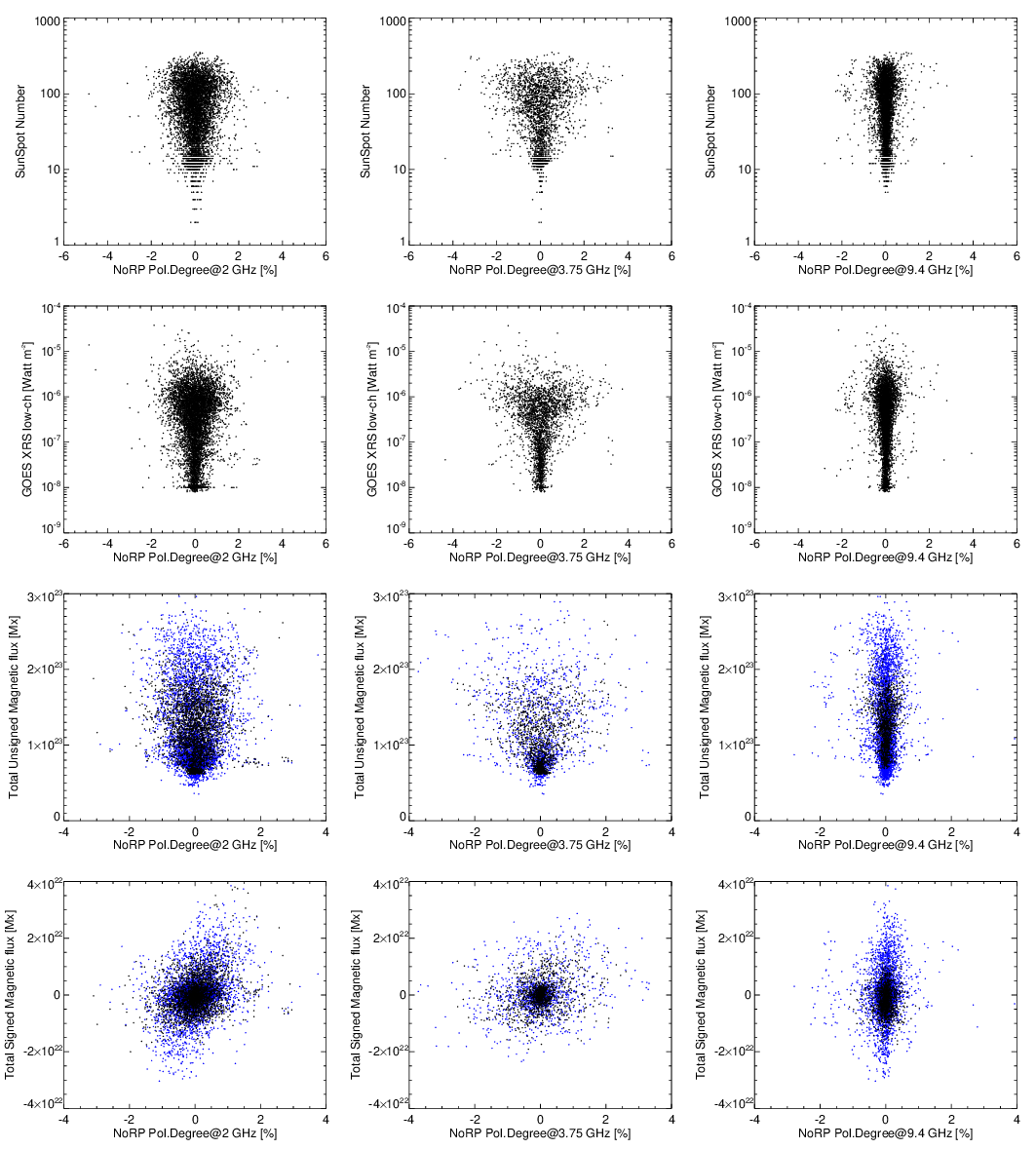}
    \caption{The scatter plots between polarization degrees and solar indices. From left, the circular polarization degrees with 2, 3.75, and 9.4 GHz. From top, Sunspot Number, GOES X-ray flux, Total unsigned magnetic flux, and Total signed magnetic flux.}
    \label{fig:poldeg}
\end{figure*}

Figure  \ref{fig:poldeg} shows scatter plots between the circular-polarization degrees at 2, 3.75, and 9.4 GHz and the solar indices from 1994 to 2020. The relations between sunspot number and X-ray flux (1st and 2nd row in Figure \ref{fig:poldeg}) show that the variance of circular-polarization degrees increases above the values (sunspot: 10; X-ray flux: $10^{-7}$ Watt m$^{-2}$)). This is similar to the relationship between the microwave fluxes and these indices. The variation in the degree of circular polarization at 3.75 GHz is larger than that at 2 and 9.4 GHz. It shows that the contribution from gyroresonance emission with 3.75 GHz is larger than the others, and it is consistent with the peak of the gyroresonance emission spectra located between 3 -- 6 GHz in solar cases \citep{1994SoPh..152..167S}.

Even when the sunspot number and X-ray flux become large, many data points locate around 0 \% of the circular-polarization degrees. When we observe only one active region on the solar disk, a circularly polarized signal can be detected, as shown in Figure \ref{fig:noaa10960}. However, when there are several active regions in each hemisphere, which is the usual condition near the solar maximum, the magnetic polarity of the preceding sunspots in the northern hemisphere is opposite to that in the southern hemisphere, following Hale's law. Thus, the circular-polarization signals from the preceding sunspots were canceled by the sunspots in the other hemisphere. Hence, there are many data points around 0 \% polarization degrees even when the solar activity is high.

The 3rd row of Figure \ref{fig:poldeg}  shows the relationship between the degree of circular polarization and the total unsigned magnetic flux. This differs from the relationship between sunspot number and X-ray flux. The variance in the circular-polarization degree increased with increasing magnetic flux, even when the enhancement in the total unsigned magnetic flux was small. This suggests that even magnetic poles, for which the magnetic flux is insufficient to form a sunspot, affect the degree of polarization in the total microwave flux.

The 4th row of Figure  \ref{fig:poldeg} shows the relationship between the degree of circular polarization and the total signed magnetic flux. Before creating the plot, we predicted that we would see a positive relationship trend because the magnetic flux imbalance would be related to the circular polarization considering the gyroresonance emission mechanism. However, the prediction is incorrect, as shown in the plots. The correlation coefficients of the relationships are less than 0.3. As mentioned previously, the lack of correlation is caused by the active regions in both hemispheres because the polarization signals from one hemisphere are canceled out by those from the other. This indicates that the summed-up polarization signals strongly depend on the magnetic field distributions of the Sun and stars.

\section{Discussion}

\subsection{What do the relationships between the solar microwave fluxes and solar indices mean?}

\begin{figure}
    \centering
    \includegraphics[scale=0.9]{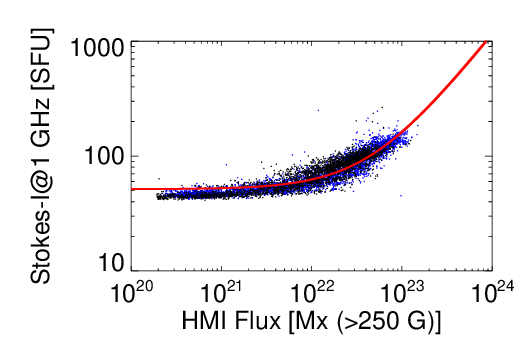}
    \caption{The relations between the 1GHz fluxes and total unsigned magnetic flux ($>$250 Gauss)}
    \label{fig:RadioMag250}
\end{figure}

As mentioned in the previous section, the relationships between the solar microwave fluxes and other activity indices can be classified into several types. First, the relationship with the microwave fluxes at any frequencies fits well with a linear functions. For example, the relationship between solar microwave flux and sunspot number (Figure \ref{fig:RadioSunspot}). To investigate these relationships, we estimated the total unsigned magnetic flux from the pixels where the magnetic field strength was over 250 Gauss. This value roughly indicates the total magnetic flux only in the active regions. Figure \ref{fig:RadioMag250} shows the relationship between the total magnetic fluxes of the active regions and the 1 GHz fluxes. Although the relationship between the 1 GHz fluxes and the total magnetic fluxes from the entire solar disk fits well with a power-law function, the relation depicted in Figure \ref{fig:RadioMag250} fits well with a linear function. This is similar to the relationship among the X-ray fluxes, sunspots, and Ca plage fractions. This suggests that the components from the active regions are dominant in the indices, in which the relationships with microwaves fit well with a linear function. This suggestion is natural because sunspot numbers and Ca II plage fractions directly indicate active regions. The X-ray flux also strongly depends on the volume of the corona above the active regions, which is also natural. When the solar activity decreases and the contribution from active regions to microwave fluxes is not dominant, the sunspot number, Ca II plage fraction, and X-ray flux become zero or negligibly small. However, even when no active region exists on the solar disk, the contribution of the thermal Bremsstrahlung from the radio photosphere dominates, and the flux converges to a constant. The constant value is determined by the average physical temperature of the radio photosphere and does not have solar cycle dependency based on observational data for over 70 years 
\citep{2017ApJ...848...62S, 2023GSDJ...10..114S}.

The other group includes the relations of microwave fluxes with the total unsigned magnetic flux or EUV flux and shows the microwave frequency dependencies. In detail, the relationship between the 1 and 2 GHz fluxes with the total unsigned magnetic fluxes and EUV fluxes shows a good fit with a power-law function. The power-law relations start breaking down around 3.75 GHz, and the scatter plots between the 9.4 GHz fluxes and these indices can be fitted well with a linear function. The transition with the observed frequency can be explained by the frequency, which depends on the opacity.

The optical depth of Bremsstrahlung rapidly increases with decreasing the observing frequency because its dependence is $f^{-2}$ in the observing frequency range of the NoRP. Therefore, the thermal Bremsstrahlung from the hot, dense plasma in the corona is optically thick below 3 GHz \citep{1999SoPh..190..309W}. As a result, increases in the 1 and 2 GHz fluxes follow that of the total unsigned magnetic flux, and the relationship is maintained even at lower flux levels. Would this claim be inconsistent with the relationship between the microwaves and X-ray fluxes? This does not matter when considering the dependence and sensitivity of plasma temperatures. The temperature response of the XRS ranges from 4 to 40 MK  \citep{2015SoPh..290.2733A}. Conversely, fluxes of a few GHz can be explained by the emission from the 1 to 3 MK plasma observed in EUV telescopes \citep{2001ApJ...561..396Z}. The X-ray flux obtained with XRS indicates only very hot ($>$4 MK) plasma that appears only with evolved active regions. In other words,  the emission from such plasma is not dominant in the microwave fluxes. Hence, the relations of the 1 and 2 GHz fluxes with the total unsigned magnetic flux differ from those with the X-ray flux obtained with XRS, even when the coronal emission strongly depends on the magnetic field. We can also understand that the 1 and 2 GHz fluxes follow the EUV flux well because the EUV emission comes mainly from plasma at such temperatures.

Above 3 GHz, especially the 9.4 GHz flux in this study, coronal structures cannot become optically thick media, and the contribution of coronal plasma cannot be dominated. In this case, the radio photosphere of the frequencies is located at the upper chromosphere  \citep[$T \sim 2 \times 10^{4}$ K for 3.75 GHz, $\sim1 \times 10^{4}$ K for 9.4 GHz:][]{1991ApJ...370..779Z}, and the 9.4 GHz flux is constructed mainly from the thermal Bremsstrahlung from radio photosphere, and thermal Bremsstrahlung from optically thin mediums in the corona that is a small contribution. In the case of the 3.75 GHz flux, the gyroresonance emission also contributes. Hence, the 3.75 GHz flux rather than the 9.4 GHz flux follows the total unsigned magnetic flux well. Given the thermal Bremsstrahlung from the radio photosphere located at the upper chromosphere dominants for 9.4 GHz and 3.75 GHz, we can easily understand that the behavior of the Mg II index, which indicates activities in the chromosphere, is similar to that of these fluxes.

Interestingly, the 3.75 and 9.4 GHz fluxes and the Mg II index converge to fixed values even when the total unsigned magnetic flux decreases. This was caused by stable emissions from the chromosphere, as mentioned earlier. In other words, this suggests that the emission from the chromosphere decouples from the magnetic field variations when the magnetic field strength is low. The sunspot number and Ca plage fraction can reach the lower limit of the measurements because they are estimated by ``counting'' using thresholds. By contrast, the observation of 3.75 and 9.4 GHz fluxes and the Mg II index do not reach such a restriction; what does maintain the minimum values of 3.75 and 9.4 GHz fluxes and the Mg II index? This may be related to the local/small-scale dynamo with convection near the surface; however, further studies are required. In recent solar-stellar relation studies \citep{2022ApJS..262...46T, 2023ApJ...945..147N}, the lowest values of the solar indices at the solar minimum are assumed to be the basal values, and the solar indices that are subtracted from the basal values are used for studies. However, we did not use the subtracted values in this study because we believed that the basal values would indicate the nonvariable components of the chromosphere during solar cycles, as mentioned earlier. For your reference, we carried on some subtracted plots shown in Appendix A.

\subsection{Knowledge transfer from solar microwave observations to stellar microwave observations}

\begin{table*}
\centering
\caption{The stellar data for Figure \ref{fig:RadioMagStellar} . The data of $\epsilon$ Eri are cited from \cite{2020ApJ...904..138S}. The data of EK Dra are cited from \cite{2017A&A...599A.127F} except for magnetic field strengths $B$.  The magnetic field strengths of these stars are cited from \cite{2020A&A...635A.142K}. Total Unsigned Magnetic Fluxes (TUMF) are calculated from the magnetic field strengths and radiuses assuming we observe the half hemisphere of the stars. }
\label{tab:stardata}
\begin{tabular}{ccccccccc}
\hline \hline
Name & Spec. Type & dis.  & Age & Rad                & $<$B$>$          & TUMF & 2-4 GHz &8-12 GHz\\
           &                   & (pc) & (Gyr)& (R$_{\odot}$) &(Gauss) &  ($10^{24}$ Mx) &  ($\mu$Jy) & ($\mu$Jy)  \\
\hline
$\epsilon$ Eri & K2V & 3.2 & 0.2 $\sim$ 0.8 & 0.74 & 186 &3.1& 29 & 68 \\
EK Dra & G1.5V & 34.0 & 0.1 & 0.95 & 1400 & 41 &- & 593 \\
\hline
\end{tabular}
\end{table*}

\begin{figure*}
    \centering
    \includegraphics[scale=0.8]{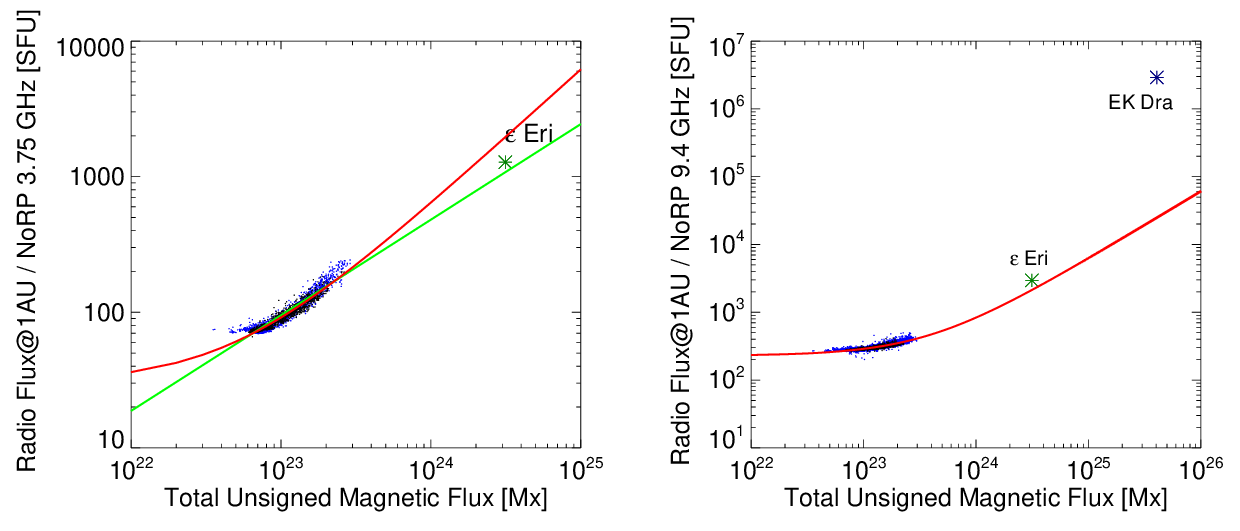}
    \caption{The relations between the 3.75/9.4 GHz fluxes and total unsigned magnetic fluxes of the solar and stellar data . The black/blue dots and red/green lines are the same as the lower panels in Figure \ref{fig:RadioMag}. The asterisks indicate the stellar data shown in Table \ref{tab:stardata}.}
    \label{fig:RadioMagStellar}
\end{figure*}

The final goal of this study was to apply the knowledge of the relationship between solar microwave data and other solar activity indices to stellar studies. As mentioned in the previous section, stellar microwave fluxes would be useful for predicting the total unsigned magnetic and EUV fluxes of stars if the relationships derived from solar data could be applied to stellar data. Hence, we must first confirm whether the solar relations can be applied to solar-type stars.

Because the microwave fluxes from solar-type stars are not sufficiently strong, stars useful for confirmation are scarce. Nevertheless, there have been some reports on the successful detection of stellar microwaves, as mentioned in Section \ref{sec:intro}. Conversely, the mean magnetic field strengths of some of these stars have been measured from the Stokes-I spectra of magnetic-sensitive lines \citep{2017A&A...599A.127F, 2015AN....336..258L}. To verify that we can apply the relationship obtained from the solar data to other data, we selected solar-type stars in which both microwave and magnetic fluxes were measured (Table \ref{tab:stardata}) and plotted the data of the stars and the Sun (Figure \ref{fig:RadioMagStellar}). The horizontal axis of the plot indicates the total unsigned magnetic fluxes of the Sun and such stars, and the vertical axis indicates the microwave fluxes of the Sun and stars, assuming that we observed the stars at a distance of 1 AU from the stars. The red and green lines show the lines extrapolated from the solar data. The data points of $\epsilon$  Eri are close to the extrapolated line, but the others are far from the lines. It is natural to consider the microwave emission mechanism. $\epsilon$  Eri is a star whose microwave emission is confirmed as thermal emission \citep{2020ApJ...904..138S}. Hence, it would be understandable that the datapoint of $\epsilon$ Eri should be located near the extrapolated lines.

Conversely, EK Dra is  the younger and more active star, and stellar flares occur frequently. In such a situation, there are nonthermal and high-energy electrons accelerated by stellar flares in their stellar corona quasi-stably, and the electrons can strongly emit microwaves via the gyrosynchrotron emission mechanism \citep{2017A&A...599A.127F}. Because it is speculated that the microwave emission from these stars originates from nonthermal electrons, the data points in Figure \ref{fig:RadioMagStellar}  are far from the extrapolated lines.

Therefore, our results suggest that we need to verify whether stellar microwave emission originates from thermal plasma to apply the relationship between solar microwaves and other activity indices presented in this paper to stellar microwave data. We try to predict the EUV flux from $\epsilon$ Eri using the solar relation because the stellar microwave is confirmed as thermal emission. Using the values shown in Table 1 and 2, we can predict the EUV flux of the star to $2.2 \times 10^{11}$ photons cm$^{-2}$ s$^{-1}$ for 26 -- 34 nm band and  $5.0 \times 10^{11}$ photons cm$^{-2}$ s$^{-1}$ for 0.1 -- 50 nm band of SEM/SOHO. These values were approximately five to seven times larger than the solar EUV flux at the solar maxima.

Only one sample is available for confirmation with the observing frequency range of NoRP, and this is not sufficient to argue for its application. We need to observe more solar-type stars within a few GHz, confirm that the emission comes from thermal plasma, and verify the application of solar radiation to stellar microwaves.

Circular polarization from stellar thermal emissions in the microwave range has not yet been detected in any star. However, based on the time variation of solar circular polarization with microwaves shown in this study, we can predict the application of stellar circular-polarization data. Recently, stellar superflares and eruptions have been detected on solar-type stars with an energy of more than  $10^{33}$ erg \citep[e.g.,][]{2012Natur.485..478M, 2022NatAs...6..241N}.  Because the estimated energy of such phenomena is large compared with that of solar phenomena, large starspots, whose size is similar to the radius of the star, appear on the stellar surface. Is a large monolithic starspot present on a star surface? Are numerous spots of both polarities creating a large dark region on the stellar surface? Investigations of structures on stellar surfaces have been conducted, and dark regions and their evolution have been revealed from the time variation of the visible light flux during a few rotation periods of the star \citep[e.g.,][]{2012A&A...543A.146F, 2017ApJ...846...99M, 2020ApJ...891..103N, 2020ApJ...901...14B}. However, answering this question is difficult because no magnetic information was available from their observations. The Zeeman-Doppler imaging (ZDI) technique is a powerful tool for mapping the distribution of magnetic fields on stellar surfaces \citep{1989A&A...225..456S,2014A&A...569A..79J}.  However, the spatial resolution of the magnetic field map obtained using the ZDI was insufficient.

The microwave circular-polarization degree is helpful for solving this problem, even when the stellar surface cannot be spatially resolved with the wavelength. As per the solar data shown in this study, when the dark region on the stellar surface is constructed from a large monolithic sunspot, the time variation of the microwave flux and its circular-polarization degree during its rotation period would be similar to those displayed in Figure \ref{fig:noaa10960}, and we can see a significant variation in both the microwave flux and circular-polarization degree. When the dark region is constructed from numerous spots with both polarities, the microwave flux increases; however, the variation in the degree of circular polarization is small because these are not correlated, as shown in Figure  \ref{fig:poldeg}. Thus, we can obtain information on the magnetic field distribution from the microwave data without spatial resolution. Before the analysis, when we obtain microwave spectra at the activity minima, we can estimate the peak frequency due to gyroresonance emission by subtracting the minimum microwave spectrum and can predict the average magnetic field strength above the starspots from the peak frequency. Nevertheless, to analyze the aforementioned parameters and cases , we need a highly sensitive telescope and high-accuracy measurements of polarization degree ($<$1 \%), which have not yet been realized by using the current instruments for most solar-type stars. Therefore, these investigations would be conducted over the next decade using the SKA and ngVLA. 
Although the realistic highest sensitivity with the current interferometers around 2 GHz is in the order of 10 $\rm \mu$Jy, the predicted sensitivity of such next-generation interferometers is  in the order of 0.1 $\rm \mu$Jy. Based on these values, we made  a rough estimate of the number of stars detectable with such instruments from the astrometric catalogue obtained with the Hipparcos mission \citep{2007A&A...474..653V}. The details of the estimation are described in Appendix B. The result of the estimation shows that we can detect thermal emission from several tens of stars at stellar minima and a few hundred stars at stellar maxima using SKA and ngVLA. The number of the stars would be sufficient to verify our results.

\appendix
\section{The scatter plots from the subtracted microwave fluxes}

\begin{figure*}
   \centering
    \includegraphics[scale=0.9]{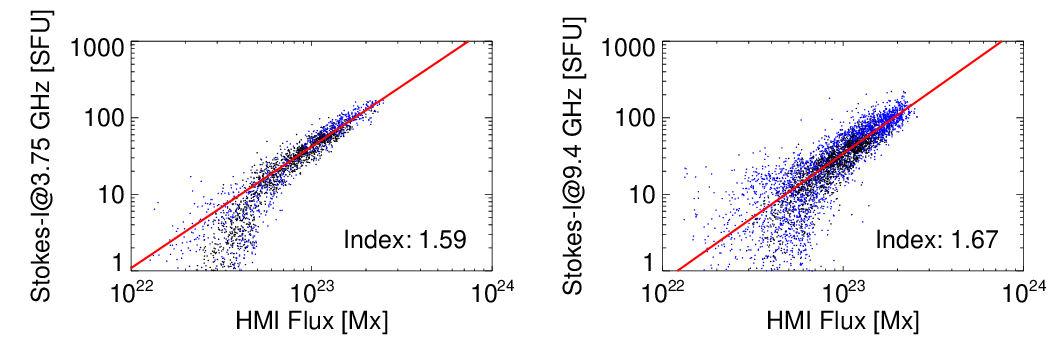}
    \caption{The relations between the solar microwave fluxes subtracted the basal fluxes and total unsigned magnetic flux.  Red lines indicate the results of fitting with a power-law function. 
                  The fitting ranges are magnetic flux: $>8 \times 10^{22}$ Mx, Microwave: $>10$ SFU}
    \label{fig:RadioMagsubB}

    \includegraphics[scale=0.9]{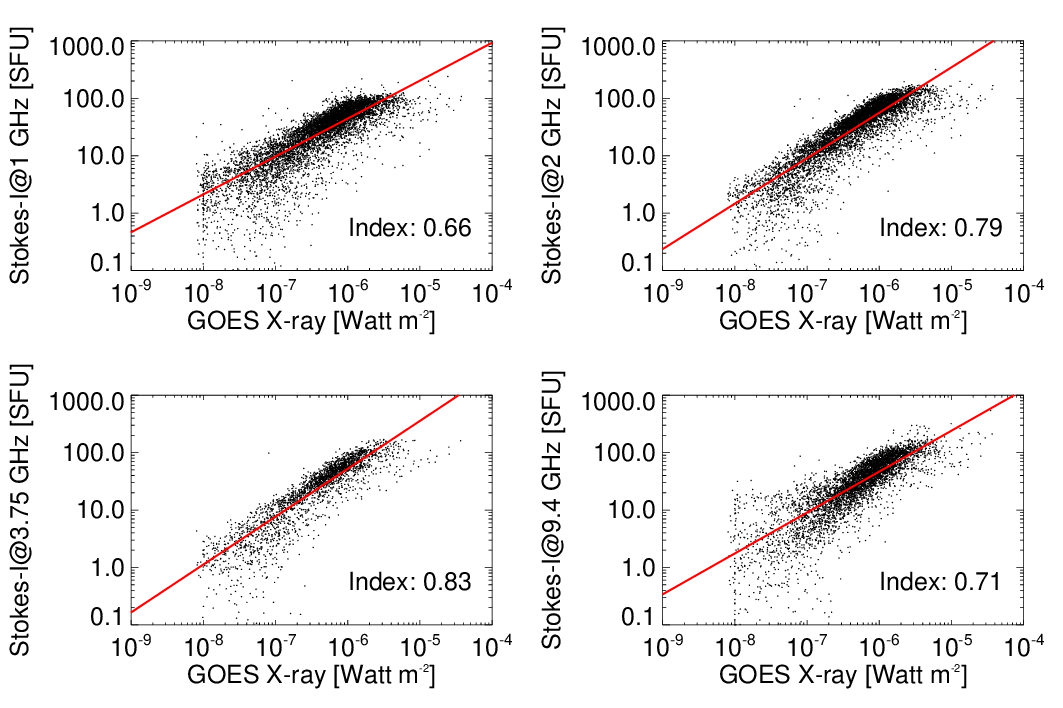}
    \caption{The relations between the solar microwave subtracted the basal fluxes and X-ray fluxes. Red lines indicate the results of fitting with a power-law function. 
                  The fitting ranges are X-ray: $10^{-8} - 10^{-5}$ Watt m$^{-2}$, Microwave: $> 1.0$ SFU}
    \label{fig:RadioXraysubB}
\end{figure*}

As mentioned in Section 4.2, the lowest values of the solar indices at the solar minimum indicate the lowest state of the Sun. Since these values are essential for understanding the lowest state of solar-type stars, we have not subtracted the lowest values (basal fluxes) to make the plots in this paper. However, as shown in Figure 2 of \cite{2022ApJS..262...46T}, the plots without the basal flux are good for presenting the relationship. To compare the figures in \cite{2022ApJS..262...46T}, we created the scatter plot between the subtracted solar microwave fluxes at 3.75, 9.4 GHz and the total unsigned magnetic flux (Figure \ref{fig:RadioMagsubB}) and the plot  between the subtracted solar microwave fluxes at all NoRP's observing frequencies and the X-ray flux (Figure \ref{fig:RadioXraysubB}). To generate the plots, we defined the "a" parameters of X-ray in Table  \ref{tab:fitResLin}  as the basal flux for microwave and defined $4 \times 10^{22}$ Mx as the basal flux for total magnetic flux.

\section{The number of the detectable solar-type stars}

To estimate the number of detectable stars, we obtained the Hipparcos astrometric data \citep{2007A&A...474..653V} of the F8 - K2 stars from the VizieR catalog access tool \citep{2000A&AS..143...23O}. We used the B-V color obtained with the Hipparcos mission and the B-V color-temperature relation \citep{2000AJ....120.1072S} to estimate the effective temperatures, and then created the HR diagrams (the upper left panel in Figure 18). In the temperature estimation, we did not use the terms later than the 5th term in equation 2 of  \cite{2000AJ....120.1072S}. From the HR diagrams, we selected the stars below the red-dashed lines as solar-type stars. From the absolute luminosity and the effective temperature of the Sun and the stars, we estimated the radius of the stars (the upper right panel in Figure 18).

\begin{figure*}
   \centering
    \includegraphics[scale=0.9]{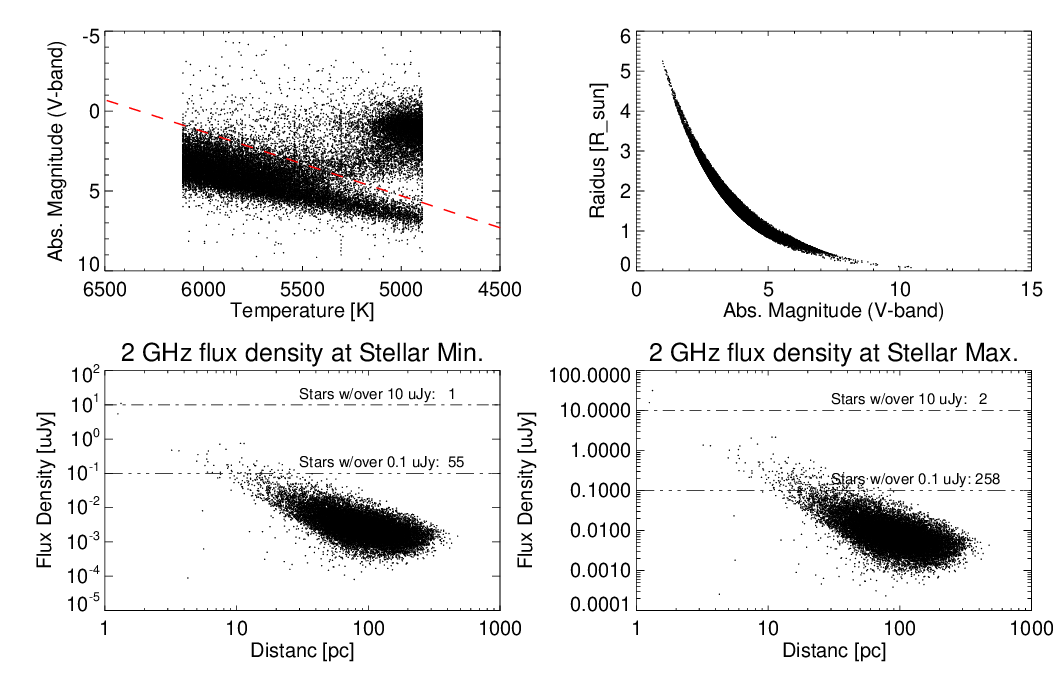}
    \caption{The details of the number of the detectable solar-type stars}
    \label{fig:SelStar}
 \end{figure*}

\cite{2017ApJ...848...62S}  show that the solar fluxes with 2 GHz at solar minima and maxima are about 50 SFU and 150 SFU, respectively. We assumed that the spatially averaged brightness temperatures of the stars are the same as the values that can explain these solar fluxes. From the spatially averaged brightness temperatures, the stellar radius, and the distance to the stars, we estimated the flux density at 2 GHz from the stars (the lower panels in the Figure \ref{fig:SelStar}). The left panel shows the flux densities at the stellar minimum, and the right panel shows at the stellar maximum. The dashed dot and dashed triple dot lines indicate the 10 and 0.1 $\rm \mu$Jy levels, respectively. The numbers in the plots indicate the number of stars whose flux densities at 2 GHz exceed these levels.

\begin{acknowledgments}
The authors would like to express their gratitude to the people who participated in the operation of the solar radio telescopes at the Nobeyama Radio Polarimeter (NoRP). The authors also thank Dr. Yang Liu (W. W. Hansen Experimental Physics Laboratory, Stanford University), Dr. Mark Weber (Institute of Environmental Physics, University of Bremen), Dr. Don Woodraska (Laboratory for Atmospheric and Space Physics, University of Colorado), and Dr. Shin Toriumi (JAXA/ISAS). Their comments on the observational data were extremely useful for this study. The NoRP is operated by the Solar Science Observatory, a branch of the National Astronomical Observatory of Japan, and its observational data are scientifically verified by the consortium for NoRP scientific operations. The SOlar and Heliospheric Observatory (SOHO) is a project of international cooperation between ESA and NASA. The Solar Dynamic Observatory (SDO) is the first mission to be launched for NASA's Living With a Star Program. The data analysis for this study was carried out on the Multi-wavelength Data Analysis System (MDAS) and the Solar Data Archive System (SDAS) operated by the Astronomy Data Center, National Astronomical Observatory of Japan. This research has made use of the VizieR catalogue access tool, CDS, Strasbourg, France (DOI : 10.26093/cds/vizier). The original description  of the VizieR service was published in \cite{2000A&AS..143...23O}. This study was supported by the Japan Society for the Promotion of Science (JSPS) KAKENHI, Grant Numbers 21J00316 (K.N.) and 22K03710 (K.W.).
\end{acknowledgments}

\facility{NoRP, GOES, SDO, SOHO, HIPPARCOS}

\bibliography{NoRP_SV.bib}{}
\bibliographystyle{aasjournal}

\end{document}